\documentclass[aps,prd,twocolumn,groupedaddress,pdftex]{revtex4-1}
\usepackage{graphicx}  
\usepackage{url}
\usepackage{bm}        
\usepackage{amssymb,amsmath}   
\usepackage{verbatim}
\usepackage[T1]{fontenc}
\usepackage[usenames,dvipsnames]{xcolor}
\usepackage{graphics}
\usepackage[colorlinks=true,
    linkcolor=red!60!black,
    citecolor=blue!80!black,
    urlcolor=blue!80!black]{hyperref}
\usepackage{float}
\usepackage{cleveref}               

\usepackage{ifpdf}

\usepackage{amsmath}

\usepackage{mathtools}

\usepackage{dcolumn}
\usepackage{tikz}
\usetikzlibrary{shapes.geometric, arrows}
\usepackage{upgreek}
\usepackage{setspace}
\usepackage{enumitem}
\usepackage{array,multirow,bigdelim}
\newcommand{\ord}[1]{{\scriptscriptstyle (#1)}}

\newcommand{\beq}{\begin{equation}}
\newcommand{\eeq}{\end{equation}}
\newcommand{\beqq}{\begin{equation*}}
\newcommand{\eeqq}{\end{equation*}}
\newcommand\beqa{\begin{eqnarray}}
\newcommand\eeqa{\end{eqnarray}}
\newcommand\beqaa{\begin{eqnarray*}}
\newcommand\eeqaa{\end{eqnarray*}}
\newcommand\bea{\begin{array}}
\newcommand\eea{\end{array}}

\begin{document}

\title{Gravitational Metric of a Star}
\author{Poul H. Damgaard$^1$, Hojin Lee $^2$, Kanghoon Lee$^3$ and Tabasum Rahnuma$^3$\\ }
\affiliation{\smallskip $^1$Niels Bohr International Academy\\ Niels Bohr Institute, University of Copenhagen\\
Blegdamsvej 17, DK-2100 Copenhagen \O, Denmark \\
$^2$School of Physics, Korea Institute for Advanced Study, 85 Hoegi-ro, Dongdaemun-Gu, Seoul 02455, Korea\\
$^3$Quantum Universe Center, Korea Institute for Advanced Study, 85 Hoegi-ro, Dongdaemun-Gu, Seoul 02455, Korea\smallskip}

\begin{abstract}
Solving the classical equations of motion in general relativity recursively, we consider the
metric of a spatially localized and stationary source of matter. Having in mind a star of general composition, we characterize it by means 
of its infinite set of mass and current multipoles. Specializing to de Donder gauge we set up the recursive equations that produce
the metric outside the star to any desired order in perturbation theory, expanded both in Newton's constant and in the order
of multipoles. Up to second post-Minkowskian order we express the result to any order in the multipole expansion in terms of
generalized (tensor) bubble integrals in momentum space and a corresponding simple expansion in inverse distances. 
In a special corner of
the space of multipoles we recover the Kerr black hole solution to the given order. By tweaking just slightly the multipoles
away from the Kerr limit the metric will describe stars that are Kerr-like and yet are not black holes.
A subtlety with respect to the gauge ambiguity of de Donder gauge is also pointed out. 
\end{abstract}
\maketitle




\section{Introduction}

What is the space-time metric outside a star? It is an elementary property of Newtonian gravity that a spherically symmetric mass distribution leads to a gravitational potential
that at any distance $r$ from the center is equivalent to that of a point-like source of mass equal to the total mass inside a sphere of that radius. In general relativity, an
analogous statement is Birkhoff's theorem, which implies that a spherically symmetric vacuum solution of the Einstein equations is that of a Schwarzschild black hole, 
$i.e.$, as if sourced
by a point-like mass at the origin. But real stars are not perfectly spherically symmetric, and it is thus of both theoretical and experimental interest to determine the effects
of deviations from spherical symmetry. Again, Newtonian gravity gives us a suitable tool with which to analyze the issue: the multipole expansion of the mass distribution.
In this approach, the arbitrary mass distribution inside a given finite volume is parametrized by its infinite sequence of multipole moments, with higher multipole moments
corresponding to weaker and weaker contributions to the gravitational field at large distances away from the star. The multipole expansion in that case is simply that of an
expansion in inverse powers of radius $r$.

In general relativity, the situation is more involved. We can still consider a distribution of matter sources inside a finite three-volume and ask for the gravitational metric outside
that finite volume. Intuition might tell us that at very large distances, the vacuum solution could simplify. Should it perhaps approach 
that of a black hole, and if so, how could this be borne out in detail?
A multipole expansion is again a suitable starting point, but there are issues that should be confronted first. One concerns the possibility of gravitational radiation. Even if the mass distribution of the star is
not perfectly axisymmetric, but deviates just slightly from this symmetry, we can consider the metric as effectively time-independent in an adiabatic domain of time scales, where we can ignore the effect of gravitational radiation. In this paper we refrain from going into the technical details on how to define such an adiabatic time interval during which it is sensible to ignore gravitational radiation. Instead, we appeal to known astrophysical sources, as well as associated astrophysical time scales over which gravitational radiation remains negligible. This may be so even though there can still be clear observational signatures specific to Einstein's general theory of relativity and the deviations from Newtonian physics. As an example we can have in
mind our own sun which definitely causes measurable general relativistic effects without measurable gravitational radiation being one of them.
These considerations lead us to focus on the case of {\em stationary} matter sources and corresponding time-independent space-metric, the vacuum
solutions of the Einstein equations outside a finite region of three-space which contain all the matter sources. The study of metrics by means of the multipole expansion in
general relativity
dates back to Geroch \cite{Geroch:1970cd} and Hansen \cite{Hansen:1974zz}, and in particular to the comprehensive treatment of Thorne 
\cite{Thorne:1980ru}, who introduced the notion of Asymptotically Cartesian and Mass-Centered (ACMC) coordinates. By restricting
ourselves to stationary sources, we allow for uniform rotations. Such stationary
matter distributions inside a finite volume we will define to be our star. In a particular limit, and as we will clarify below, we recover the multipole expansion of a Kerr black hole.

The most important difference with a Newtonian multipole problem is obviously the non-linearity of general relativity. In Newtonian gravity, the gravitational field outside a star can be
described by its infinite set of multipole moments, and hence by a simple expansion in terms of powers of $1/r$, each term corresponding to a given multipole moment. The
gravitational field outside a Newtonian star is thus in immediate one-to-one correspondence with the multipole moments, and we can conversely read off
each multipole moment as the coefficient of a given $1/r^n$ fall-off at infinity. It follows from solving the associated Poisson equation,
which is exact. In general relativity, the gravitational interactions imply that this is so only at the linearized level. 
Beyond, the various moments that characterize the mass distribution and currents of the stationary sources are still in one-to-one correspondence with the gravitational metric away from such a bounded source, but this is filtered by the gravitational self-interactions.
The $n$'th order term of the expansion with fall-off $1/r^n$ will then, in the weak-field approximation, have an additional 
infinite expansion in powers of Newton's constant $G$. This makes the multipole formalism of general relativity a double expansion: in $1/r^n$ and in powers of $G$. It is
customary to refer to this as a multipolar post-Minkowskian (MPM) expansion. Its use by far surpasses the immediate goals of this paper, which is restricted to the
application to stationary space-times. Away from the stationary limit the multipole expansion is a crucial analytical tool with which to describe gravitational radiation from localized sources of matter, and there is a large and impressive literature on the subject (see, $e.g.$, \cite{Gursel:1983nkl,Thorne:1984mz,Blanchet:1985sp,Blanchet:1986dk,Damour:1990gj,Blanchet:1992br,Damour:1994pk,Blanchet:1995fr,Blanchet:1997ji,Zschocke:2019zdm} for a partial list).
Finding the gravitational metric of a compact object such as a neutron star of course also has a long history, dating back to the seminal work of Tolman, Oppenheimer, and Volkoff \cite{Tolman:1939jz,Oppenheimer:1939ne}. The aim was initially to find the gravitational metric outside a spherical source of matter corresponding
to a given equation of state. There is also a large literature on the external gravitational field of rotating stars and how to map from a given equation of state onto a set of multipole moments of more general distributions of matter with cylindrical symmetry, see, $e.g.$, refs. 
\cite{Hartle:1968si,Ryan:1995wh,Ryan:1997hg,Laarakkers:1997hb,Pappas:2012ns,Yagi:2013awa}.

We shall here revisit the multipole expansion of general relativity in the light of the recursive approach of ref. \cite{Damgaard:2024fqj}. Briefly stated, this allows us
to express in a compact manner the iterative nature of the perturbative solution to Einstein's field equations, giving each new order in $G$ in terms of integrals over lower-order
expressions. The integrations arise due to the solutions of field equations being most conveniently obtained by Fourier-transforming to momentum space. For the stationary space-times,
we consider in this paper those integrations that are over three-dimensional space only, as in \cite{Damgaard:2024fqj}. This considerably simplifies computations, and the required
integrals are all of a simple kind that we label as generalized bubble integrals (tensor integrals of rational form corresponding to having massless propagators in quantum field
theory language). These bubble integrals arise simply when we introduce the inverse Laplacian in momentum space, and they are loop integrals only in that sense. At intermediate stages, we need to regularize the integrals, and it is
natural to choose dimensional regularization, which is ideally suited for this problem. 
One of the computational obstacles in the early approach to the multipole expansion was indeed the use of the more cumbersome Hadamard regularization, which in
addition requires non-trivial amendments beyond the first leading orders (the use of dimensional regularization for the multipole problem in general relativity 
was advocated earlier in refs. \cite{Blanchet:2003gy,Blanchet:2005tk}). We note already here the use of both terminology and tools from relativistic quantum field theory. This paper is
indeed one more step towards reformulating analytical solutions to the Einstein field equations by means of a recursive quantum field theoretic technique. 
Great strides have already been
made based on the quantum field theoretical amplitude-based approach \cite{Bjerrum-Bohr:2018xdl,Cheung:2018wkq,Kosower:2018adc,Bern:2019nnu,Cristofoli:2019neg,DiVecchia:2020ymx,Bjerrum-Bohr:2021vuf,Bjerrum-Bohr:2019kec,Bern:2021dqo,Herrmann:2021lqe,Damgaard:2021ipf,Brandhuber:2023hhy,Kim:2024svw}. It considers the scattering of point-like particles that interact only gravitationally. Extracting the
classical terms of such computations is a known procedure, and the common lore in recent years has been that it is far simpler to first consider the full quantum field theoretical
amplitude problem of the post-Minkowskian expansion, and subsequently extract only the classical terms, than it is to solve the classical equations of motion directly. We 
challenge such a position in this paper by considering a problem with no simple and immediate 
description in terms of the scattering of point-like massive particles. 
An alternative formulation preceding the iterative approach of ref. \cite{Damgaard:2024fqj} that we also pursue here is the off-shell recursive method
of refs. \cite{Cho:2021nim,Lee:2022aiu,Lee:2023zuu}. It contains similar iterative structures originating in the perturbiner formalism, but it also includes all quantum corrections and is thus closely related to the full amplitude-based formalisms.
One quantum field theoretical
approach that is much closer to our perturbative set-up is the worldline formalism for classical general relativity, which precisely selects the classical
pieces to each order in $G$ \cite{Goldberger:2004jt,Kalin:2020mvi,Mogull:2020sak,Jakobsen:2021smu,Dlapa:2021npj,Jakobsen:2022psy,Kalin:2022hph,Jakobsen:2023ndj}. These worldline formulations can include dissipative effects due to gravitational radiation as well, and are in fact in one-to-one 
correspondence \cite{Damgaard:2023vnx} with the KMOC-prescription for amplitudes \cite{Kosower:2018adc} which also includes dissapative back-reaction from the gravitational field.

The worldline formalisms are, however, still rooted in point-like descriptions of massive 
particles that interact gravitationally. The link between the multipole expansion and new effective couplings to the worldline action is part of the program of describing
extended objects in general relativity through what is known as the gravitational skeleton method, whereby the worldline becomes dressed with higher-order couplings to
the gravitational field \cite{Goldberger:2004jt}. 
Instead, we here proceed directly to the solution of the classical equations of motion, now using to advantage the recent progress in
integration techniques from scattering amplitude computations. This bypasses the need for effective field theory ideas, matching
conditions, and other mappings between the fundamental degrees of freedom of classical Einstein gravity and those of the 
effective field theory. 
Everything is determined by classical physics, classical equations of motion, and the coupling constant $G$. The only assumption needed is that a perturbative expansion around flat Minkowski space is meaningful so that the multipole expansion is applicable.

We choose the post-Minkowskian multipole expansion of stationary matter sources to illustrate the advantages of this recursive formalism. The resulting solution is algebraic in nature, with the solution being provided by tensor integrals of increasing rank. Inevitably, the actual solution becomes of rapidly growing complexity with this rank. Fourier transformation back to three-space yields a corresponding complicated
tensor structures of Cartesian coordinates, and the expansion rapidly clogs up due to the numerous different tensors that are allowed and which will appear. 
Such is the perturbative multipole formalism
since each new order is of higher rank, and the non-linear nature of the field equations leads to a mixing of these tensor structures at higher orders in $G$.
It is also not our aim here to provide detailed expressions to very high order in the
post-Minkowskian expansion because they are not by themselves particularly illuminating. We shall nevertheless present what we believe is
the first complete 2nd order post-Minkowskian expression
for stationary space-times based on mass multipoles up to quadrupoles, and current multipoles up to dipole order, in order to illustrate the relative ease with which we obtain these results. Pushing this to higher orders can be done straightforwardly. It has been argued that such highly accurate multipole expansions for celestial mechanics in our solar system will soon be required to match the observational accuracy of sub-micro-arcsecond astrometry \cite{Zschocke:2019zdm,Zschocke:2025umr}.

\section{The Post-Minkowskian Multipole Expansion}\label{defsec}

From Newtonian gravity (or electrostatics) we are familiar with the multipole expansion, typically expressed in terms of spherical harmonics $Y_{\ell m}(\theta,\phi)$ or 
associated Legendre polynomials
$P_{\ell}^m(\cos\theta)$. For our purposes, the alternative formulation in terms of symmetric trace-free (STF) tensors (see ref.~\cite{Thorne:1980ru} for a clear exposition of the properties of such STF tensors) is simpler to deal with. We shall
briefly review the relation between the two formulations below. STF tensors arise very naturally when taking Cartesian coordinate derivatives of $1/r$ and are thus very 
naturally linked to a multipole expansion. In addition, such STF tensors satisfy a number of beautiful relations that simplify analytical expressions. As for spherical harmonics,
which are scalar functions on the sphere, STF tensors appear because of their interpretation in terms of representations of the rotation group in three dimensions. In
order to make the paper self-contained, we begin this section with an overview of the relation between these two formulations, and simultaneously, we introduce our
choice of variables, conventions, and notation. A number of very useful relations among STF tensors are relegated to Appendix A.


It is well-known that a convenient choice for gravitational perturbation theory is an expansion around flat space of the
gothic (or Landau-Lifshitz) tensor density,
\begin{equation}
    \mathfrak{g}^{\mu \nu}(x^\mu) \equiv \sqrt{-g} \, g^{\mu \nu} = \eta^{\mu \nu} - h^{\mu\nu} ,
\end{equation}
which selects harmonic coordinates $x^{\mu}$ by the gauge condition $\partial_{\nu}(\mathfrak{g}^{\mu \nu} ) = 0$. Perturbation
theory is set up by expanding the deviation from Minkowski space as an infinite series in $G$,
\beq
h^{\mu\nu} ~=~ \sum_{n=1}^{\infty} G^n h_{(n)}^{\mu \nu} ~, \label{hexp}
\eeq
and solving the equations of motion order by order in $G$.
Then each term $ h_{(n)}^{\mu \nu} (t, x,y,z) $ of the series in harmonic coordinates admits a finite multipolar expansion associated with the $ O(3) $ group of rotations of 
the spatial coordinates (which leave invariant  $ r \equiv ((x^1)^2 + (x^2)^2 + (x^3)^2)^{\frac{1}{2}} $ and $ t \equiv x^0$), $i.e.$,
\begin{equation}\label{stfmoments}
  h_{(n)}^{\mu \nu} (x^\mu) = \sum_{l=0}^{\infty} h_{{(n)},L}^{\mu \nu} (r, t) \, \hat{n}^L (\theta, \phi),
\end{equation}
Using a standard notation, $ L $ denotes a multi-index 
$ i_1 i_2 \dots i_l $, and thus $ n^L := n^{i_1} n^{i_2} \dots n^{i_l} $ with $ n^i \equiv x^i / r $ ($i = 1, 2, 3 $). A hat on Cartesian tensors indicates that we take the symmetric trace-free
(STF) part. Thus, $ \hat{n}^L $ denotes the STF part of $ n^L $ and as an
example $\hat{n}^{ij} = n^in^j - \delta^{ij}/3$. We refer to Appendix \ref{appendixA} for general STF tensor expressions and useful relations among them.
The sum appearing on the right-hand side of 
Eq.~\eqref{stfmoments} is equivalent to a finite expansion in spherical harmonics $ Y_l^m(\theta, \phi)$ as explained in detail in
ref. \cite{Thorne:1980ru}.

As mentioned in the Introduction, we define our star by a stationary and spatially localized source of matter. The resulting spacetime metric in the vacuum outside the star will then 
be time-independent. Such a definition includes isolated and non-evolving compact bodies, such as non-radiating stars and also black holes. Far from the spatially localized sources, one can expand the metric in terms of multipole moments in order to describe the gravitational field generated by them. These multipole moments characterize the spatial distribution of mass and currents in a gauge-invariant (coordinate-invariant) way.
At each order in $G$, there is an infinite set of STF tensors representing mass $M$ and current $S$ multipoles: 
$$\{M_L , S_L\}, \quad L = i_1 \cdots i_l.$$
After solving the Einstein equations with these sources, the multipoles thus determine the entire external field and they are essential for precision modeling of gravitational interactions. In the asymptotic zone of a stationary source, this multipole expansion is time-independent. 

The time-independent mass and current multipoles are defined by moments of the energy-momentum tensor
\cite{Damour:1990gj,Zschocke:2019zdm},
\begin{enumerate}
    \item \textbf{Mass Multipole Moments:}
\begin{equation}
M^{L} = \int d^3x\, \left(T^{00}(\boldsymbol{x}) + T^{ii}(\boldsymbol{x})\right)\, \hat{x}^{L} 
\label{MLdef}\end{equation}
where $ \hat{x}^{L} \equiv x^{\langle i_1} x^{i_2} \cdots x^{i_l \rangle} $ denotes the STF part of the multi-index product, and the angle brackets represent the STF projection. Note that we should distiguish $\hat{x}^{L}$ and $x^{L}$, which $x^{L}$ is the convetional tensor product $x^{L} = x^{i_1} x^{i_2} \cdots x^{i_l}$.
\item \textbf{Current Multipole Moments:}
\begin{equation}\label{currentmultipole}
  S^{L} = \int d^3x\, x^{\langle L-1} \, \epsilon^{i\rangle}_{j k} x^j T^{0k}(\boldsymbol{x})\,,
\end{equation}
where $x^{L-1} = x^{i_{1}} x^{i_{2}} \cdots x^{i_{l-1}}$.
\end{enumerate}
We note that a classic paper describing the approximate metric of a rotating stationary matter distribution is that of Hartle and Thorne \cite{Hartle:1968si}. This predates the
multipole expansion literature and shows how to obtain an approximate metric for, for instance, a neutron star by a slowly rotated Schwarzschild solution, truncated at leading terms.

\section{The Recursive Solution}

Here we briefly review the iterative method that was used to compute the Schwarzschild metric to all orders in ref. \cite{Damgaard:2024fqj}. It boils down to systematizing the perturbative solution to the Einstein equations of motion in de Donder gauge based on the expansion of Eq.~\eqref{hexp}. We have already provided the first step in the previous section. Restricting ourselves to stationary solutions, we here consider time-independent sources of matter that give rise to an energy-momentum tensor $T_{\mu\nu}(\boldsymbol{x})$ which, as indicated, is only a function of the spatial coordinates $\boldsymbol{x}$. We have collected a number of useful identities involving STF tensors and their relation to Fourier-transformed variables in Appendix B. Using results from there, and to leading order in $G$, the equations of motion then reduce to Eq.~\eqref{h1eq} and with the multipole expanded solution of Eq.~\eqref{h1sol}.

It is convenient to introduce the Fourier transforms of $h^{\mu\nu}$ which we call currents. Thus, corresponding to the expansion of Eq.~\eqref{hexp}, we also introduce an ordered sum
\begin{equation}
  J^{\mu\nu}(\boldsymbol{k})
  =
  \sum_{n=1}^{\infty} G^n J_{(n)}^{\mu \nu}(\boldsymbol{k}) ~,
\label{Jexp}\end{equation}
where 
\begin{equation}
  J_{(n)}^{\mu \nu}(\boldsymbol{k})
  \equiv
  \int_{\boldsymbol{x}} h_{(n)}^{\mu\nu}(\boldsymbol{x})e^{-i \boldsymbol{k} \cdot \boldsymbol{x}} \,.
\label{Jdef}\end{equation}
For simplicity of notation we define $J_{(n)|\boldsymbol{k}}^{\mu \nu} \equiv J_{(n)}^{\mu \nu}(\boldsymbol{k})$. 
Using the terminology of ref. \cite{Damgaard:2024fqj} we call $n$ the rank of the expansion. For rank $1$ we have linearized gravity, and
the currents are simply the Fourier transforms of Eq.~\eqref{h1sol}. Higher-order terms derive from the gravitational self-interactions 
that are included in the Einstein tensor. Expanding the equations of motion order by order in $G$ and making use of the de Donder gauge condition, 
we collect terms in the Landau-Lifshitz pseudo-tensor 
\beq
\tau_{LL}^{\mu \nu} = \sum_{n=1}^{\infty} G^{n+1}\tau^{\mu \nu}_{LL(n)} ~.
\eeq
Matching powers of $G$ this defines a hierarchy of equations. The expansion of the Landau-Lifshitz pseudo-tensor starts at one order higher
than the expansion of $h^{\mu\nu}$ itself, and the equations become recursive. Each new order of $h^{\mu\nu}_{(n)}$ is given in terms of combinations
of previously computed terms of up to and including those of rank $(n-1)$. The usefulness of the currents of Eq.~\eqref{Jexp} derives simply from
the need to solve the Green function problem for the $n$th term, thus making it convenient to first go into momentum space, perform the needed 
convolutions, and then
transform back to real space. The intermediate step of Fourier-space convolutions is the analog of loop integrations in the amplitude approach.
Beyond the first few orders, the expansion becomes algebraically unwieldy. This problem can be solved by introducing the auxiliary set of metric tensor variables described in ref. \cite{Damgaard:2024fqj}. It bypasses at intermediate steps the inversion of $\mathfrak{g}^{\mu \nu}$ through a doubling of metric variables by the introduction of a new tensor density $\tilde{\mathfrak{g}}_{\mu \nu}$ which is only identified as the inverse of $\mathfrak{g}^{\mu \nu}$ ($i.e.$, $\mathfrak{g}^{\mu \nu}\tilde{\mathfrak{g}}_{\nu \rho} = \delta^{\mu}_{\rho}$) at the end. Such a simplified method becomes essentially mandatory to introduce at high orders in $G$ but it is not needed here.

For the 2PM calculations of this paper, we need only the first non-trivial terms of the    Landau-Lifshitz pseudo-tensor $\tau_{LL(1)}^{\mu \nu}$. The Einstein equation at the 2PM order is
\begin{equation}
\begin{split}
  \Box h^{\mu \nu}_{(2)}(\boldsymbol{x})
  &=
  -\tau_{LL(1)}^{\mu \nu}\,,
\label{2PMEoM}\end{split}
\end{equation}
where, if we denote $h = h_{\rho}{}^{\rho}$,
\begin{widetext}
\begin{equation}
\begin{aligned}
  \tau_{LL(1)}^{\mu \nu}
  &=
   2 \partial^{\rho} h^{\sigma(\mu} \partial_{\sigma} h_{\rho}{}^{\nu)}
  -2 \partial_{\rho}\big(h_{\sigma}{}^{(\mu} \partial^{\nu)} h^{\rho \sigma}\big)
  +2 h_{\rho}{}^{(\mu} \square h^{\nu) \rho}+\frac{1}{2} \partial^{\mu} h^{\rho \sigma} \partial^{\nu} h_{\rho \sigma}-\partial_{\rho}\left(h^{\rho \sigma} \partial_{\sigma} h^{\mu \nu}\right) 
  -\frac{1}{4} \partial^{\mu} h \partial^{\nu} h
  \\&\quad
  -\frac{1}{2} h^{\mu \nu} \square h
  +\eta^{\mu \nu}\left[
      \frac{1}{8} \partial^{\rho} h \partial_{\rho} h
    + \frac{1}{4} h \square h
    +\frac{1}{4} \partial^{\rho} h^{\sigma \tau} \partial_{\rho} h_{\sigma \tau}-\frac{1}{2} h^{\rho \sigma} \square h_{\rho \sigma}
    +\frac{1}{2} \partial^{\tau} h^{\rho \sigma} \partial_{\rho} h_{\tau \sigma}
    +h^{\rho \sigma} \partial_{\rho} \partial^{\tau} h_{\sigma \tau}
  \right]\,.
\end{aligned}\label{2PMPseudoEMTensor}
\end{equation}
\end{widetext}
which we will use in the next section. We point out the important fact that from this stage and further  towards higher orders in $G$, all dynamics resides entirely in the
gravitational self-interactions of general relativity. The higher-order terms are thus universal (invariant functions of the particular values of the infinite set of multipole moments that arise from the energy-momentum tensor) and reflecting precisely how general relativity glues these different multipole moments together based on the principle of general covariance.

\section{Truncated Multipole Expansions}\label{sectionIV}
Having set up the recursive relations, we are now ready to apply them to the multipole expansion. 
In contradistinction to the Schwarzschild case \cite{Damgaard:2024fqj} the recursion is now double: both in order of multipoles and in order of $G$. The recursion relations expand in 
powers of $G$, given what can be viewed as initial data in
terms of 1PM multipoles. This is analogous to the Schwarzschild solution, where the initial condition is 
provided by a single point-like matter source at the origin. That Schwarzschild case corresponds to a time-independent metric of non-vanishing $h^{00}$, all other components of $h$ vanishing at that initial order. For the general multipole expansion, the source is rather characterized by an infinite set of physical parameters that describe the Taylor expansion of the metric at the linearized level. In practice, one will truncate at a given order in multipoles and then generate the post-Minkowskian expansion in $G$ based on this finite number of initial values. This double expansion rapidly clogs up due to the numerous different tensor structures that result from the couplings between different multipoles, together with spatial indices. In order not to lose sight of the underlying simple recursive structures in the post-Minkowskian expansion, it is therefore illustrative to begin with an example to 2PM order, truncated to quadrupole mass multipole order and leading current multipole order. As we shall see, this already produces new results, and it will then become clear how it generalizes to arbitrary orders.

\subsection{Rank 1}\label{1pmrec}
We recall that we label the power $n$ of the coupling constant $G$ as the rank. Rank 1 is what in quantum field theory language, would be labeled tree level, and  $\mathrm{h}^{\mu \nu}_{(1)}$ is the solution to the classical equations of motion to order $G$. This is the starting point of the iteration and we have the simple relation Eq.~\eqref{h1eq} of linearized gravity. The solutions \eqref{h1sol} of this equation give our initial conditions for the recursion relations.

Hence, from \eqref{Jdef} we get the rank-1 currents up to the quadrupole moments
\begin{equation}\label{initialcond}
\begin{split}
  J^{00}_{(1)|\boldsymbol{k}} 
  &=
  \int_{\boldsymbol{x}} \left( \frac{4M}{r} + \frac{6M_{ab} }{r^3} \hat{n}_{ab} \right)e^{-i \boldsymbol{k} \cdot \boldsymbol{x}}\,,
  \\
  &=
  16 \pi \left( \frac{M}{|\boldsymbol{ k}|^2} -\frac{M_{ab}\hat{ k}^{ab}}{2 |\boldsymbol{k}|^{2}} \right)\,,
  \\
  J^{0i}_{(1)|\boldsymbol{ k}}
  &=
  \int_{\boldsymbol{x}} \left( -2 \epsilon_{iab} n^a \frac{S^b}{r^2}\right)e^{-i \boldsymbol{ k} \cdot \boldsymbol{x}}
  \\
  &= i\epsilon_{iab}\frac{8 \pi S^b k^a}{|\boldsymbol{ k}|^{2}}  ,
\end{split}
\end{equation}
and in addition $\mathcal{J}^{ij}_{(1)|\boldsymbol{ k}} = 0$.

\subsection{Rank 2}

From Einstein equation at 2PM order \eqref{2PMEoM}, we first consider the contributions to the time components $h^{00}_{(2)}$ under the stationary condition and the harmonic gauge:
\begin{equation}
\begin{aligned}
  &\nabla^2 h_{(2)}^{00} 
  =
  \frac{7}{8} \partial_i h_{(1)}^{00} \partial_i h_{(1)}^{00} +\frac{5}{4} h_{(1)}^{00} \partial^2 h_{(1)}^{00} 
  \\&\quad
  -\frac{1}{2} \partial_i h_{(1)}^{0j} \partial_i h_{(1)}^{0j} -\frac{3}{2} \partial_i h_{(1)}^{0j} \partial_j h_{(1)}^{0i}
    - h_{(1)}^{0i} \partial^2 h_{(1)}^{0i}\,,
\end{aligned}\label{2PMEOM}
\end{equation}
which by iteration gives us the currents. To help displaying how the recursion works, we split up $h_{(2)}^{00}$ into two parts: $h^{00[a]}_{(2)}$, which depends only on $h^{00}_{(1)}$, and $h_{(2)}^{00[b]}$, which depends only on $h^{0i}_{(1)}$,
\begin{equation}
  h_{(2)}^{00} = h_{(2)}^{00[a]} + h_{(2)}^{00[b]}\,,
\label{}\end{equation}
where $h_{(2)}^{00[a]}$ and $h_{(2)}^{00[b]}$ satisfy
\begin{equation}
\begin{aligned}
  \nabla^{2}h_{(2)}^{00[a]}
  &=
  \frac{7}{8} \partial_i h_{(1)}^{00} \partial_i h_{(1)}^{00} +\frac{5}{4} h_{(1)}^{00} \partial^2 h_{(1)}^{00}\,,
  \\
  \nabla^{2}h_{(2)}^{00[b]}
  &=
  -\frac{1}{2} \partial_i h_{(1)}^{0j} \partial_i h_{(1)}^{0j} 
  -\frac{3}{2} \partial_i h_{(1)}^{0j} \partial_j h_{(1)}^{0i}
  \\&\quad
  -h_{(1)}^{0i} \partial^2 h_{(1)}^{0i}\,.
\end{aligned}\label{}
\end{equation}

Let us consider first the iterative contributions from $J^{00[a]}_{(2)|-\boldsymbol{k}_1}$, which is associated with $\partial^{2}h_{(2)}^{00[a]}$
\begin{equation}\label{2PMMassM} \begin{split}
  J^{00[a]}_{(2)|-\boldsymbol{k}_1} 
  &=
  \frac{1}{|\boldsymbol{k}_1|^2} \int_{\boldsymbol{k_2}} 
  \left[ \frac{5}{4}|\boldsymbol{k}_2|^2 {-} \frac{7}{8} \boldsymbol{k}_{12}{\cdot}\boldsymbol{k}_2 \right]
  \\&\quad\times
  J^{00}_{(1)|-\boldsymbol{k}_{12}}J^{00}_{(1)|\boldsymbol{k}_2}\,.
\end{split}
\end{equation} 
Using the bubble integral formula in Eq.~\eqref{2PMBubble}, we have
\begin{equation}
\begin{aligned}
  \frac{1}{\pi^{2}}J^{00[a]}_{(2)|-\boldsymbol{ k}}
  &=
      \frac{14M^2}{|\boldsymbol k|}
  {+} \frac{7 M_{ab}M_{ab} |\boldsymbol k|^3}{40} 
  {+} \frac{3 M^{2}_{ab} \hat{ k}^{ab} |\boldsymbol k|}{8}
  \\&\quad
  {-} \frac{21 M M_{ab}\hat{ k}^{ab}}{4 |\boldsymbol k|}
  {+} \frac{21 M_{ab}M_{cd}}{256 |\boldsymbol k|}\hat{ k}^{abcd} .
\end{aligned}\label{2PMcurrent}
\end{equation}
We next find the contributions from the three terms involving the $h_{(1)}^{0i}$ components, $viz.$, those arising
from the current multipoles. We label them by superscript $[b]$:
\begin{equation}\label{2PMcmultipole}
\begin{aligned}
  |\boldsymbol{k}_1|^2J^{00[b]}_{(2)|-\boldsymbol{k}_1}
  &{=}
  \int_{\boldsymbol{k_2}} \bigg[
    \bigg( \frac{\boldsymbol{k}_{12} {\cdot} \boldsymbol{k}_2}{2} {-} |\boldsymbol{k}_2|^2\bigg) \delta^{ij} {+} \frac{3\boldsymbol{k}_{2}^i \boldsymbol{k}_{12}^j}{2}\bigg]
  \\&\qquad
  \times J^{0i}_{(1)|-\boldsymbol{k}_{12}}J^{0j}_{(1)|\boldsymbol{k}_2}
\end{aligned}
\end{equation}
Similarly, upon substitution of the corresponding 1PM currents from \eqref{initialcond}, this gives
\begin{equation}\label{current2PM}
\begin{split}
  &|\boldsymbol{k}_1|^2J^{00[b]}_{(2)|-\boldsymbol{k}_1}
  \\ 
  &{=}
  \int_{\boldsymbol{k_2}} \bigg[~
    \frac{32\pi^2\boldsymbol{k}_{12}{\cdot} \boldsymbol{k}_2}{|\boldsymbol{k}_2|^2 |\boldsymbol{k}_{12}|^2} 
    \big( \boldsymbol{k}_{2}{\cdot}\boldsymbol{k}_{12} |\boldsymbol{S}|^2 
          {-} (\boldsymbol{k}_{12} {\cdot} \boldsymbol{S} ) (\boldsymbol{k}_{2}{\cdot}\boldsymbol{S}) \big)
  \\&\qquad\qquad
    - \frac{96\pi^2}{|\boldsymbol{k}_2|^2 |\boldsymbol{k}_{12}|^2} \big(\boldsymbol{k}_{2} \cdot (\boldsymbol{k}_{12} \times \boldsymbol S)\big)^{2}
  \bigg]\,,
\end{split}
\end{equation}
where $\boldsymbol{k}_{2} \cdot (\boldsymbol{k}_{12} \times \boldsymbol S) = \epsilon_{ijk} k^{i}_{2} k^{j}_{12} S^{k}$. Evaluating the integrals we obtain
\begin{equation}\label{c2PM}
\begin{split}
  J^{00[b]}_{(2)|\boldsymbol{ k}}
  =
    \frac{5\pi^2|\boldsymbol{S}|^2 |\boldsymbol k|}{3}
  - \frac{7\pi^2S^i S^j }{4|\boldsymbol k|}\hat{ k}^{ij} \,.
\end{split}
\end{equation}
The total 2PM contribution is thus $J^{00}_{(2)| k}  = J^{00[a]}_{(2)| k} + J^{00[b]}_{(2)| k}$ with the result
\begin{equation}\label{Joi2PM}
\begin{split}
  J^{00}_{(2)|\boldsymbol{ k}}
  &=
  \frac{14\pi^2  M^2}{|\boldsymbol k|} + \frac{7 \pi^2 M_{ab}M_{ab} |\boldsymbol k|^3}{40} + \frac{5\pi^2|\boldsymbol{S}|^2 |\boldsymbol k|}{3}
  \\&~
  +\frac{3\pi^{2}}{4}\bigg[ \frac{M^{2}_{ab} |\boldsymbol k|^{2}}{2|\boldsymbol k|}{-}\frac{7 M M_{ab}}{|\boldsymbol k|} {-}\frac{7S_{a}S_{b}}{3|\boldsymbol k|} \bigg]\hat{ k}^{ab}
  \\&~
  +\frac{ 21 \pi^2 M_{ab}M_{cd}}{ 256 |\boldsymbol k|}\hat{ k}^{abcd}\,.
\end{split}
\end{equation}

We next move on to find the currents for $h^{0i}$ to order $G^2$. From the equations of motion to this order,
\begin{equation}\label{2PMeom}
  \nabla^2 h_{(2)}^{0i}
  =
    \frac{1}{2} h_{(1)}^{0i} \partial^2 h_{(1)}^{00}
  + \partial_j h_{(1)}^{0i} \partial_j h_{(1)}^{00}
  - \partial_i h_{(1)}^{0j} \partial_j h_{(1)}^{00} \,,\end{equation}
and the corresponding recursion is 
\begin{equation}
\begin{aligned}
  J^{0i}_{(2)|-\boldsymbol{k}_{1}}\!
  &=
  \frac{1}{|\boldsymbol{k}_{1}|^{2}} \int_{\boldsymbol{k}_{2}}
  \Big[
    \big(\tfrac{1}{2}|\boldsymbol{k}_{2}|^{2}{-}\boldsymbol{k}_{12}{\cdot}\boldsymbol{k}_{2}\big)\delta^{ij}
    + k^{i}_{12} k^{j}_{2}
  \Big]  
  \\&\qquad
  \times J^{0j}_{(1)|-\boldsymbol{k}_{12}} J^{00}_{(1)|\boldsymbol{k}_{2}}\,,
\end{aligned}\label{}
\end{equation}
which for the currents gives
\begin{equation}\label{J0icomp}
\begin{split}
  J^{0i}_{(2)|-\boldsymbol{ k}}
  &=
  i \pi^2 |\boldsymbol k| \bigg[ \frac{2M\epsilon_{ijk}}{|\boldsymbol k|^{2}}
  {-} \frac{ \epsilon_{ijl}M_{lk}}{2}
  {-} \frac{M_{il}\epsilon_{ljk}}{8} \bigg] S^{j}  k^{k}\,,
  \\&\quad
  +\frac{\epsilon_{i ab} M_{cd}}{16 |\boldsymbol k|} S^{a}  k^{bcd}\,.
\end{split}
\end{equation}
We finally turn to the $h^{ij}$-components at 2PM order. While not present at order $G$, these terms 
are generated by iteration through the equations of motion:
\begin{equation}
\begin{split}
  &\nabla^2 h_{\ord{2}}^{ij}
  = \partial_k h_{\ord{1}}^{0i} \partial_k h_{\ord{1}}^{0j}
  {-} 2\partial^k h_{\ord{1}}^{0(i} \partial^{j)} h_{\ord{1}}^{0k}
  {-} \tfrac{1}{4} \partial_i h_{\ord{1}}^{00} \partial_j h_{\ord{1}}^{00}
  \\&\quad
  + \partial_i h_{\ord{1}}^{0k} \partial_j h_{\ord{1}}^{0k}
  + \tfrac{\delta^{ij}}{8} \Big[\partial_k h_{\ord{1}}^{00} \partial_k h_{\ord{1}}^{00}
  -4\partial_k h_{\ord{1}}^{0\ell} \partial_k h_{\ord{1}}^{0\ell}
  \\&\quad
  + 4\partial_k h_{\ord{1}}^{0l} \partial_l h_{\ord{1}}^{0k}
  + 2h_{\ord{1}}^{00} \nabla^2 h_{\ord{1}}^{00}
  - 8h_{\ord{1}}^{0k} \nabla^2 h_{\ord{1}}^{0k}\Big]\,.
\label{hij}\end{split} 
\end{equation}
Deriving the recursion by taking the Fourier trnasform, we get the corresponding currents by solving the recursion
\begin{equation}
\begin{aligned}
  &{-}\pi^{-2}J^{ij}_{(2)|\boldsymbol{ k}}
  \\
  &=\,
      M^2 A^{ij}_{1,1}
  {+} \frac{ M (\boldsymbol k{\cdot}M {\cdot} \boldsymbol k)}{16|\boldsymbol{ k}|} A^{ij}_{5,7}
  {+} \frac{M}{8 |\boldsymbol{ k}|} B^{ij}
  {-} \frac{M M^{ij}|\boldsymbol{ k}|}{8}
  \\&\quad
  {-} \frac{(\boldsymbol k{\cdot} M {\cdot} \boldsymbol k)^{2} }{512 |\boldsymbol{ k}|} A^{ij}_{5,1}
  {+} \frac{(\boldsymbol k{\cdot} M {\cdot} \boldsymbol k)}{128 |\boldsymbol{ k}|} B^{ij}
  {-} \frac{(\boldsymbol k{\cdot}M{\cdot}M{\cdot}\boldsymbol k)}{128|\boldsymbol{ k}|^{-2}}A^{ij}_{4,\frac{13}{3}}
  \\&\quad
  + \frac{|\boldsymbol{ k}|}{192}\bigg[\frac{7}{2}(\boldsymbol k{\cdot}M{\cdot}\boldsymbol k) M^{ij} 
    - 5 (M{\cdot}\boldsymbol k)^{i} (M{\cdot}\boldsymbol k)^{j}\bigg]
  \\&\quad
  {+} \frac{(M{\cdot}\boldsymbol{ k})_{k} l}{192} B^{ijk}
  {-} \frac{5l^{2} M_{ab} M_{ab}}{768} A^{ij}_{1,1}
  {-} \frac{l^{3}}{192} M^2_{ij}
  \\&\quad
  {+}
  \frac{(\boldsymbol{S}{\times}\boldsymbol{ k})_{i}(\boldsymbol{S}{\times}\boldsymbol{ k})_{j}}{2|\boldsymbol{ k}|}\,
  {+} \frac{2(\boldsymbol{S}{\cdot}\boldsymbol{ k})^{2}}{16} A^{ij}_{1,\frac{1}{3}}
  {+} \frac{|\boldsymbol{S}|^{2}|\boldsymbol{ k}|^{2}}{16} A^{ij}_{11,3}
  \\&\quad
  {+}\frac{5}{8}\bigg[|\boldsymbol{ k}| S^{i}S^{j} 
  -\frac{2\boldsymbol{ k}{\cdot}\boldsymbol{S}}{|\boldsymbol{ k}|}  k^{(i}S^{j)} \bigg]\,.
\end{aligned}\label{}
\end{equation}
where $\boldsymbol k{\cdot} M {\cdot} \boldsymbol k =  k^{i} M_{ij}  k^{j}$, $(M{\cdot}\boldsymbol k)_{i} = M_{ij}  k^{j}$ and
\begin{equation}
\begin{aligned}
  A^{ij}_{\alpha,\beta}
  &=
  \frac{\alpha  k_1^i  k_1^j}{ k^3} - \frac{\beta \delta^{ij} }{ k}\,,
  \\
  B^{ijk}
  &=
   k^{j} M^{ik}+  k^{i} M^{jk}\,,
  \qquad 
  B^{ij}
  =
  B^{ijk}  k_{k} \,.
\end{aligned}\label{}
\end{equation}

\subsection{Stationary gravitational metric at 2PM order}

The iterative solution has provided us with all stationary 2PM currents that include mass multipoles and the 
leading current multipole corresponding to angular momentum. The only remaining task in order to extract the 
metric is to Fourier transform back to space coordinates. Before listing those expressions, we here provide
a few details on how to solve the required integrals. The same method used to derive, $e.g.$, Eq.~\eqref{FourierFormula}.

Using this, we can now provide the corresponding expressions for the metric tensor,
\begin{equation}\label{metricat2PM}
\begin{aligned}
  &h^{00}_{(2)}
  = 
  \frac{7 M^2 }{r^2} 
  + \frac{ 21 M M_{ab}}{r^4}\hat{n}_{ab} -\frac{5|\boldsymbol{S}|^2 }{ 3 r^4}
  + \frac{7 S^a S^b}{ r^4} \hat{n}^{ab}
  \\&\quad
  + \frac{63M_{ab}M_{cd}\hat{n}^{abcd}}{4\, r^{6}} + \frac{9M^{2}_{ab}\hat{n}^{ab}}{r^{6}} + \frac{21M_{ab}M_{ab}}{10\, r^{6}} ,\end{aligned} 
\end{equation}
and
\begin{equation}\label{currentcom2PM}
\begin{aligned}
  h^{0i}_{(2)}
  &=
    \frac{2M(\boldsymbol{S}{\times}\boldsymbol{n})^{i}}{r^3}
  + \frac{3\tilde{S}_{ia}M_{bc} \hat{n}^{abc}}{2r^5}
  \\&\quad
  -\frac{19\tilde{S}_{ia} (M{\cdot} \boldsymbol{n})_{a}}{10r^{5}} 
  + \frac{M_{ij}(\boldsymbol{S}{\times}\boldsymbol{n})^{j}}{2r^5}  ,
\end{aligned}
\end{equation}
where $M^{2}_{ab}= M_{ac} M_{cb}$ and $\tilde{S}_{ia} = \epsilon_{iab}S^{b}.$

The $h^{ij}$ components at order $G^2$ are considerably more lengthy. We decompose the term as
\begin{equation}
  h^{ij}_{(2)}
  = 
  \frac{1}{r^{2}} H^{ij}_{2} 
  + \frac{1}{r^{4}} H^{ij}_{4} 
  + \frac{1}{r^{6}} H^{ij}_{6}
\label{}\end{equation}
and each components are
\begin{widetext}
\begin{equation}\label{ij2pm}
\begin{aligned}
  &H^{ij}_{2} =
  M^2 \bigg[\hat{n}^{ij}+\frac{1}{3}\delta^{ij}\bigg]\,,
  \\
  &H^{ij}_{4} 
  =
  M M_{ab} \bigg[\frac{15}{2}\hat{n}^{ijab} 
  + \frac{11}{7} \delta^{ij} \hat{n}^{ab}\bigg]
  -\frac{12}{7} M M^{a(i}\hat{n}^{j)}_a
  + \bigg[\frac{9 }{2 } \hat{n}^{ijab} -\frac{13}{7}\delta^{ij} \hat{n}^{ab}\bigg]S_a S_b
  - 2 \hat{n}^{ab} \tilde{S}_{ia} \tilde{S}_{jb} 
  \\&\qquad
  + \frac{32}{7} S^a \hat{n}^{(i}_{\ a}S^{j)}- \frac{S^i S^j }{15 } 
  -|\boldsymbol{S}|^2 \left[\frac{20}{7} \hat{n}^{ij}-\frac{2}{15}\delta^{ij}\right]\,,
  \\
  &H^{ij}_{6}
  =
  M_{ab} M_{cd} \bigg[\frac{75}{4}\hat{n}^{ijabcd}{+}\frac{9\delta^{ij}}{44}\hat{n}_{abcd}\bigg]
  - \frac{6 M^2_{ij}}{35}
  - \frac{90}{11} M^{2}_{ab} \hat{n}^{ijab}
  + M_{ab} M^{ab} \bigg[\frac{11}{70}\delta^{ij} 
  - \frac{25}{84} \hat{n}^{ij} \bigg]
  \\&\qquad
  - \frac{5}{21} M^{2}_a{}^{(i} \hat{n}^{j)}{}_b
  + \frac{18}{11}\, M^{bc} M^{a(i} \hat{n}^{j)}{}_{abc}
  + \bigg[
      \frac{10}{21} M^{ab} M^{ij} 
    {-} \frac{23}{42} M^{a(i} M^{j)b} 
    {+} \frac{29}{42} \delta^{ij} M^{2}_{ab} \bigg] \hat{n}_{ab}\,.
\end{aligned}
\end{equation}
\end{widetext}

We have checked that all these expressions match with the literature
\cite{Blanchet:1992br,Blanchet:1997ji}. 

\section{Reduction to the Kerr metric}

As a further check on the expressions given above, we can consider the black hole limit where our metric should reduce to the Kerr metric in harmonic gauge \cite{Cook:1997qc,Lin:2014laa}. This is a degenerate corner of the general multipole expansion in which the current multipoles and mass multipoles are not independent but scale with one parameter, the reduced angular momentum $a = J/M$. We start by listing the standard form of harmonic gauge $h^{\mu\nu}$ for the Kerr metric,
\begin{equation} \label{h_components}
\begin{aligned}
  h^{00}
  &=
  1+\frac{R^2 \big[a^2 (3\bar{M}^2+4\bar{M}R+R^2)+(\bar{M}{+}R)^4\big]}{a^2-\bar{M}^2+R^2}
  \\&\quad
  +\frac{a^{2}z^{2}}{a^2 z^2+R^4}\,,
  \\
  h^{0i}
  &= 
  -\frac{2 a \bar{M} R^2 (\bar{M}+R)\epsilon_{ij3}x^{j}}{\left(a^2 z^2+R^4\right) \left(a^2-\bar{M}^2+R^2\right)}\,,
  \\
  h^{ij}
  &=
  \frac{\bar{M}^2 R^2}{R^4+a^2z^2}\zeta^i \zeta^j \,,
\end{aligned}
\end{equation}
where $\bar{M} = GM$ and the defining equation for $R^{2}$ is
\beq\label{Rdef}
R^2 = r^2 - a^2\left(1 - \frac{z^2}{R^2}\right)\,,
\eeq
and $r \equiv \sqrt{x^2 + y^2 + z^2}$. 
In addition, 
\begin{equation}
  \zeta^i = \left(\frac{Rx+ay}{R^2+a^2}, \frac{-ax+Ry}{R^2+a^2}, \frac{z}{R}\right) \,.
\label{}\end{equation}
We remind that we can read off the multipoles for Kerr by expanding the terms of order $G$ and $a$ in powers of $1/r$. As mentioned earlier, there are indeed no contributions to $h^{ij}$ at order $G$, but there are terms in $h^{ij}$ with both odd and even powers of $a$ at order $G^2$ and higher. This will become an important point when we make contact with our general multipole expansion, as will be discussed below.

Indeed, an important point to mention is the following. The full Kerr solution provides one compact vacuum solution to Einstein equation outside a suitable source with spin. From that closed-form solution one cannot identify the combinations of different multipole moments that are behind the compact expression for the metric. Analyzing how the Kerr metric builds up perturbatively, order by order in $G$ and spin $a$, we can obtain that map. This will allow us to trace precisely which multipole moments contribute to given orders of 
the Kerr metric expansion. Such an analysis has not been carried out before.

We now need to identify the Kerr mass and current multipoles. An axisymmetric source such as that of a Kerr black hole with a chosen symmetry axis $s^i$ implies invariance under only the subgroup $SO(2)_s\subset SO(3)$ of rotations about $s^i$. Since rank-$l$ STF tensors form irreducible $SO(3)$ representations, imposing invariance under this axial $SO(2)$ selects a one-dimensional invariant subspace within it. In such a case, the only rotationally invariant STF direction is that of the symmetry axis (the spin $s$-axis).

Therefore, at each $l$, the multipole tensor must be proportional to the unique axis-aligned STF tensor built solely from the unit axis vector,
\begin{equation}
    M_L = a_l\,\hat{s}_{L}\,.
\end{equation}
One can find the detailed proof of this statement from symmetry in Appendix \ref{fixedmultipole}. 
This implies, if we choose the symmetry axis to be in the $3$-direction, the only independent component is the all-$z$ one $M_{33\cdots 3}$, while any component carrying an $x$
or $y$ index is excluded by the symmetry (and the remaining diagonal pieces are fixed by the trace-free
condition).

By expanding \eqref{h_components} at the $ \mathcal{O}(G)$, 
\begin{align}
\begin{split}
  h^{00}_{(1)}
  &= 
  \sum_{l=0}^{\infty} \frac{4(-1)^{l} M a^{2l}}{r^{2l+1}} P_{2l}(\hat{n}\cdot\hat{s})\,,
  \qquad
  \hat{n}\cdot\hat{s} = \frac{z}{r}\,,
  \\
  &=
  4 \sum_{l=0}^{\infty} \frac{ (-1)^{l}M a^{2l}}{r^{2l+1}}\,\frac{(4l-1)!!}{2l!} \hat{n}^{L}\ \hat{s}_{L}\,,
\end{split}
\end{align}
where, in the last step, we employed the identity in Eq.~\eqref{identity} with $L = i_1,\ldots,i_{2l}$.

We now compare with the general form in Eq.~\eqref{MPMexpansion}
and by matching coefficients, find
\begin{equation}
    M_L = (-1)^l M a^{2l}\hat{s}_{L}\,.
\end{equation}

Therefore:
\begin{equation}
\begin{aligned}
  \text{Odd}  &: M_{2l+1} = 0\,,
  \\
  \text{Even} &:  M_{2l} = M(-1)^{l} a^{2l} ~.
\end{aligned}\label{}
\end{equation}
The Kerr current multipole expressions can be read off similarly. In summary, the Kerr multipoles corresponding to an axisymmetric rank-$l$ STF tensor aligned with the spin axis-$s$ can be grouped together in the complex combination  \cite{Geroch:1970cd,Hansen:1974zz}
\begin{equation}
  \mathcal{M}_l \equiv M_l + i\, S_l ~,~~~~~~ \mathcal{M}_l = M (i a)^l\,,
\label{}\end{equation}
corresponding to
\begin{equation} \label{multipolesKerr}
\begin{split}
   M_L = (-1)^{l}a^{2l} M \;\hat{s}^{L};
   \quad
   S^{aL} = (-1)^{l} M a^{2l +1} \hat{s}^{aL}\,.
\end{split}
\end{equation}
Let $s^i$ be the unit spin direction of Kerr, chosen along the 3-direction: $s^i \equiv \delta^{i3}$. We then have 
for the current dipole and the mass monopole and quadrupoles:
\begin{equation}
\begin{split}
    S^a &= Ma\,\delta^{a3}=\left( 0,0,Ma\right),\\
M_{ab}& = -Ma^2\,\hat{s}_{ab}= -Ma^2\!\left(s_a s_b-\frac{1}{3}\delta_{ab}\right).
\end{split}
\end{equation}
With the symmetry axis aligned with the $z$-axis, we thus have for the mass quadrupole
\begin{equation*}
    \begin{split}
M_{zz}=-\frac{2}{3}Ma^2,\qquad
M_{xx}=M_{yy}=+\frac{1}{3}Ma^2 ~.
    \end{split}
\end{equation*}

We begin with a comparison between $h^{00}$ of our multipole expansion and $h^{00}$ of the Kerr metric. Since we have kept mass multipoles up to and including quadrupoles but only the leading current multipole (spin), and since we now expand in the common parameter $a$ we need some care. Through iterations, we can recover all terms of order $a^2$ (arising from combinations of mass monopole-quadrupole ($MM_{ab}$) or current dipole-dipole ($S_aS_b$)) and we can also recover all terms
of order $a^4$ arising from mass quadrupole-quadrupole ($M_{ab}M_{cd}$). But there are also potentially iterative terms coming from mass monopole-hexadepole
($MM_{abcd}$) and current dipole-octupole ($S_aS_{bcd}$) combinations. Those pieces we have not computed for the 2PM order. One could imagine also terms of
order $a^4$ from further combinations like $S_aS_bS_{cd}$, etc., but such contributions cannot arise at order $G^2$ since this is the first iteration in $G$ that
hence links only two multipoles. 

Because we have truncated our multipole expansion at mass quadrupole and current dipole order, we do not have all the pieces that contribute to order $a^4$ for $h^{00}$ at 2PM order. It turns out, however, that we have more information available than what can be read off from the `exact' form of the Kerr metric in Eq.~\eqref{h_components}.
This follows from the analysis of \cite{kerrpaper}, which shows how to build up the Kerr metric from the perturbative expansions in both $g$ and $a$, given a (1PM) source for a Kerr black hole. In detail, this can be done by attaching `charges' $Q[n]$ to all multipoles so that $Q[0]$ multiplies the mass monopole, $Q[2]$ multiplies the mass quadrupole (and in general all $Q[n]$ for $n$ even multiply the mass multipoles). Likewise, $Q[1]$ multiplies the current dipole, and in general $Q[n]$ multiplies the current multipoles for $n$ odd. This way of identifying how different multipole combinations build the Kerr metric \eqref{h_components} from a Kerr source gives us the possibility to check our 2PM computations for all components of $h^{\mu\nu}$ up to order $a^4$. We list all of the components for reference in Appendix \ref{kerrdecom}. 
For instance, for the case at hand, the expansion of the 2PM Kerr Metric $h^{00}_{2\text{PM}}$ truncated at 
$\mathcal{O}(a^4)$ is given by \cite{kerrpaper},
\begin{align}\label{Kerr2pm}
\begin{split}
    &M^{-2}h^{00}_{(2)} \big|_{\mathcal O(a^4)}
    \\
    &=
    \frac{7Q[0]^2}{r^2}
    {+}\frac{7(r^2{-}3z^2)}{r^6} Q[0]Q[2] 
    {-}\frac{(4r^2{-}7z^2)}{r^6}Q[1]^2
    \\&
    {+}\frac{7(r^2{-}3z^2)^2}{4r^{10}}Q[2]^2 
    {-}\frac{(4r^4{-}33r^2z^2{+}35z^4)}{r^{10}}Q[1]Q[3]
    \\&
    {+}\frac{7(3r^4-30r^2z^2+35z^4)}{4r^{10}}Q[0]Q[4]\,.
\end{split}
\end{align}
We are now ready to compare this with the general multipole expansions of Eq.~\eqref{metricat2PM} by substituting values for mass and current multipoles for the Kerr black hole given in Eq.~\eqref{multipolesKerr}. We recover the Kerr metric for the
specific choice 
\begin{equation}
  Q[l] ~=~ a^{l} ~. 
\label{Q}\end{equation}
The coefficients to the various combinations of these $Q$'s in this expansion of the metric correspond one-to-one with the corresponding combinations of multipole moments. They show how the Kerr metric builds up perturbatively by combining those different multipoles through the recursion.

Collecting the results up to quadrupole moments $h^{00}_{(2)}\big|_{\rm Quad}$, we have
\begin{equation}\label{Kerr00match}
 \begin{aligned}
      h^{00}_{(2)}\big|_{\rm Quad}\!
  =\!
  \frac{M^2}{r^{2}}\bigg[7 {+}\frac{a^{2}(3-\! 14n_{3}^{2})}{r^{2}}
  + \frac{a^{4}(1 -\! 3 n_{3}^2)^{2}}{4r^{4}}\bigg] ~,
 \end{aligned}
\end{equation}
with $n_{3} = \frac{z}{r}$.
We can now clearly see how our general expansion reduces to the Kerr values through the match to the first four coefficients given in Eq.~\eqref{Kerr2pm} after substitution of the identification \eqref{Q}. In greater generality, assigning these $Q$-labels at the $1$PM source level of the Kerr metric allows us to identify and extract different multipole coefficients in the metric at any higher PM order and higher multipole orders. 

We now turn to a similar check on our $h^{0i}$-expressions at 2PM order. At leading order in $a$ only the current dipole and 
mass monopole, $i.e$, $MS_a$, contribute. From \eqref{currentcom2PM} we immediately obtain at order $a$:
\begin{equation}
  h^{0i}_{(2)} = -2M\epsilon^{ij3} S_3 \frac{n_j}{r^3} = -2Ma\frac{\epsilon^{ij3}x_{j}}{r^3} \,.
\label{}\end{equation}
Moving on to order $a^3$ (only odd orders of $a$ can appear due to parity) we can have combinations $M_{ab}S_c$, $MS_{abc}$, and $MM_{abc}$. 
Of these, we can only check
the terms arising from quadrupole-dipole contributions.

Then we can write, 
\begin{equation}
\begin{split}
  h^{0i}_{(2)}
  &= \,M^2 a\,\epsilon_{iab}n^a S^b
  \left[
    -\frac{2}{r^3}
    -\frac{a^2}{2r^5}\left(3n_{3}^{2}+1\right)
  \right]\,.
\end{split}
\end{equation}
Comparing with the component expressions for the Kerr metric at 2PM order listed in Appendix \ref{kerrdecom}, 
we find a perfect match to order $a^3$ from our general multipole expansion. 

We finally turn to a comparison of our $h^{ij}$-expressions in the Kerr limit. At 2PM order we can potentially 
get terms of order $a$ (from combinations
$MS_a$), order $a^2$ (from combinations $S_aS_b$ and $MM_{ab}$), order $a^3$ 
(from combinations $S_aM_{bc}$ and $MS_{abc}$), and order $a^4$ (from combinations
$M_{ab}M_{cd}$, $MM_{abcd}$, $M_{abc}S_d$, and $MS_{abcd}$). Of these, we are not comparing with any terms 
involving $M_{abc}, M_{abcd}, S_{abc}$ or
$S_{abcd}$. We now check which of the remaining kinds of combinations will appear in our iterative solution.

As we have chosen the spin axis, $s^i$ to be in the $3-$direction ($s^{i} \equiv \delta^{i3}$), we have
\begin{equation}\label{kerr2pmex}\begin{split}
  &h^{ij}_{(2)}
  = 
    \frac{M^2}{r^2}n^i n^j
  \\&
  + \frac{M^2 a^2}{r^4}\Big[
      \left(5n_{3}^{2}-11\right)n^i n^j
    -2 v^i v^j
    +8n_{3}s^{(i} n^{j)}
    \\&\qquad\qquad
    -3s^i s^j
    +3\left(1-n_{3}^{2}\right)\delta^{ij}
  \Big]
  \\&
  +\frac{3M^{2}a^{4}}{4r^{6}} \Big[
      \big(1-18n_{3}^{2}+25n_{3}^{4}\big)n^{i}n^{j}
    + 2\big(n_{3}^{2}{-}\tfrac{1}{3}\big) s^{i} s^{j}
  \\&\qquad\qquad
    + 8(n_{3}-2n_{3}^{3}) s^{(i}n^{j)}
    + 2n_{3}^{2}(1-n_{3}^{2})\,\delta^{ij}
  \Big]\,,
\end{split}
\end{equation}
where $v^{i}= \epsilon^{i}{}_{a3} n^{a}$.

This matches with the Kerr metric at mass quadrupole-quadrupole order $\mathcal{O}(a^4)$, as 
listed in Appendix \ref{kerrdecom}. 

We thus find agreement with the Kerr metric in the Kerr limit of multipoles in Eq.~\eqref{multipolesKerr} for all terms we have 
computed at 2PM order.
There are two ways in which to view these confirmations of results. 
The first is to see this as only a consistency check on two similar computations: the perturbative solution based on the Kerr source of ref. 
\cite{kerrpaper} (which is based on the same iterative formalism), and our more general multipole expansion here. These two calculations should clearly agree in the Kerr limit. Another way to consider it is to think beyond the Kerr black hole and view the deformations from the Kerr metric as representing interesting compact astrophysical objects whose metrics can resemble Kerr black holes to as high a degree of accuracy as we wish. This second viewpoint highlights the interesting scenario that a compact star can seem arbitrarily similar to a Kerr black hole, and yet it will not be a black hole. Approaching the star from a far distance, all multipoles may appear to be those of the Kerr metric, except for a few multipoles that are slightly
different. Then there will be no horizons, and the star will reveal its true identity only very close to the would-be horizons. The existence
of such a black hole mimicker should not be ruled out by fundamental principles in general relativity. Neither should it be ruled out by astrophysical 
production mechanisms that would force the stellar evolution to end at a Kerr black hole if sufficiently close to it in terms
of multipoles. It would not seem to involve careful tuning to realize such an astrophysical compact object. Examples of external
metrics of rotating compact objects that differ by arbitrarily small amounts from those of Kerr black holes
have been considered in, $e.g.$, refs. \cite{Quevedo:1986nn,Bonga:2021ouq}. It is worth stressing that a way to distinguish spinning black holes from
mimickers in gravitational  wave signals would be  to extract individual multipole moments from the data. Although impossible in the near future, and very difficult in the long run, this
would provide a quantifiable measure of the degree of confidence one has in assigning the identities of genuine black holes solutions to the binary systems. The idea put forth in this paper is that for stationary sources one can compute the full multipole expansion beyond linearized gravity to in principle arbitrarily high order in $G$, and thus be able to identify not only potential deviations from black hole metrics in the data but also pinpoint precisely which combinations of multipole moments will be impacted by such potential deviations. In practice, the distinction between black holes and very compact astrophysical objects may be almost immaterial from the point of view of gravitational wave physics, but of course of major importance for astrophysics.

Computing the Kerr metric perturbatively from scattering amplitudes and worldline formalisms is a different line of attack which assigns spin to a point-like object either at small quantized levels or for spins of classical magnitudes through effective couplings  to curvature terms \cite{Donoghue:2001qc,Barker:1975ae,Porto:2005ac,Holstein:2008sx,Damgaard:2019lfh,Bern:2020buy,Gambino:2024uge,Aoude:2022trd,Haddad:2024ebn,Cangemi:2022bew,Bjerrum-Bohr:2023jau,Ben-Shahar:2023djm}. In the amplitude-based approach this is achieved through the addition of non-minimal
couplings to higher-spin fields, and an approach similar in spirit is taken in the worldline formalisms. The link between these two constructions is through the higher-derivative non-minimal couplings, which again map to the set of multipoles of the present paper.

\subsection{Gauge redundancy}

We finally point out an important caveat with respect to the matching of our general multipole expansion with that of the Kerr metric in the Kerr limit. While we find agreement to the 2PM order with the specific combinations of multipole moments described above, there are also terms in the expansion of Eq.~\eqref{h_components} that we cannot reproduce, and we should explain why. As an illustration, consider terms proportional to $a$ at the 2PM contributions to $h^{ij}$. If we expand Eq.~\eqref{h_components} to that order, we find the term
\begin{equation}
  h^{12}_{(2)} = \frac{ M^2 a}{r^5}(y^2 - x^2) \,,
\label{gaugeterm}\end{equation}
which can only arise from a combination of mass monopole and current dipole, $MS_a$. However, as can be seen from \eqref{hij}, a term of that kind will not be generated by our recursion. This apparent inconsistency of our recursive method is resolved once it is realized that the `exact' de Donder gauge Kerr metric of Eq.~\eqref{h_components} is only one particular form of the Kerr metric in de Donder gauge, an issue explained very clearly in ref. \cite{Fromholz:2013hka} for the case of the Schwarzschild black hole. Because the harmonicity condition on coordinates is preserved under a shift of those coordinates by any harmonic function, we have the freedom to add such a 2PM residual gauge vector $\xi^i$, 
\beq
x^i \to x^i + \xi^i,
\eeq
where $\Box\xi^i = 0$. The gothic $h^{ij}$'s do not transform as tensors under such an infinitesimal transformation, but rather 
\beq
h^{ij} \to h^{ij} - \partial^i\xi^j - \partial^j\xi^i + \eta^{ij}\partial_k\xi^k ~.
\eeq
If we choose 
\beq
\xi^i = -M^2\,a \,\epsilon^{ij3}\frac{x^j}{3r^3} ~~, ~~~ i,j \in \{1,2\}\,,
\eeq
which indeed obeys $\Box\xi^i = 0$, then $h^{12}$ transforms as
\beq
  h^{12}_{(2)} \to h^{12}_{(2)} - \frac{M^2 a}{r^5}(y^2 - x^2)\,,
\eeq
and it thus cancels the order-$a$ term of Eq.~\eqref{gaugeterm}. This phenomenon was noticed in ref. \cite{kerrpaper} and it shows that the standard form of the Kerr metric in harmonic gauge Eq.~\eqref{h_components}, which nicely presents a compact non-perturbative version of the harmonic-gauge Kerr metric, achieves the compact closed-form expression by including terms that can be removed by a coordinate change while still retaining the harmonic gauge.
\par
In fact, this gauge redundancy persists to infinite order in $a$. Although we do not need it here, we note that the corresponding 2PM gauge vector to all orders in $a$ is most conveniently expressed in terms of the variable $R$, and it reads 
\beq
\xi^i = \frac{M^2}{2a^2}\left[\frac{Ra}{R^2+a^2} + \tan^{-1}\left(\frac{R}{a}\right) - \frac{\pi}{2}\right]\epsilon^{ij3}x^j \,,
\eeq
with the relation between $R$ and $r$ as given in Eq.~\eqref{Rdef}. The existence of this gauge vector implies that at 2PM order, all terms with
odd powers of $a$ are gauge artifacts within the de Donder gauge. Indeed, our recursive solution does not generate such terms.


\section{Generalization to Higher Multipoles}

We finally show how the above iteration generalizes to any number of multipoles. Since the 
${\cal O}(G^2)$ equations of motion are unchanged, the recursion relations are unaltered, but the currents that enter these relations will include more terms. Correspondingly, new convolution integrals will arise. They are fortunately all of what can be called generalized bubble integrals, the nature of which follows from the fact that we are solving the massless equations of motion for the metric fluctuations. The 2PM solution for the metric involving an arbitrary number of higher multipoles can be given entirely in terms of these generalized bubble integrals. Here we outline the solution for $h^{00}$ to order $G^2$ and including all orders in the multipoles.

We begin by writing the 1PM current multipole expansion for $h^{00}$ from Eq. \eqref{MPMexpansion}: 
\begin{equation}
\begin{split}
  J^{00}_{(1)|\boldsymbol{k}} 
  &=
  4\sum_{l=0}^{\infty}\frac{(-1)^{2l} (2l - 1)!!  }{l!} M_L\int_{\boldsymbol{x}} \frac{\hat{n}_L}{r^{l+1}} e^{-i \boldsymbol{k} \cdot \boldsymbol{x}}\,.
\end{split}
\end{equation}
A useful consequence of $ M_L $ being an STF tensor is that its contraction with any other tensor $ T_L $ satisfies $M_L T_L = M_L \hat{T}_L$ where $ \hat{T}_L $ denotes the STF projection of $ T_L $. This means that we can switch freely between hatted $\hat{\boldsymbol{k}}$ and un-hatted $\boldsymbol{k}$ variables in the above expression for the 1PM current:
\begin{equation}\label{1PMtoallorders}
\begin{split}
  J^{00}_{(1)|\pm\boldsymbol{k}}
  =
  16 \pi \sum_{l=0}^{\infty} \frac{(\pm i)^{l}}{l!} \frac{ M_L\hat{k}^{L}}{|\boldsymbol{k}|^2}\,.
\end{split}
\end{equation}
Momentum space expressions for the current multipole components follow analogously. With the help of the Fourier transform formula Eq. \eqref{FourierFormula}, the 1PM current corresponding $h^{01}_{(1)}$ can be derived from Eq. \eqref{MPMexpansion}
%
\begin{equation}
  J^{0i}_{(1)|\pm\boldsymbol{k}}
  =
  16 \pi\sum_{l=0}^{\infty} \frac{(\pm i)^l l}{(l+1)!} \epsilon_{iab}\frac{S_{bL-1}\hat{k}^{aL-1}}{|\boldsymbol{k}|^2}\,,
\end{equation}
together with $J^{ij}_{(1)|\pm\boldsymbol{k}} = 0$.

\subsection{Iterations for rank-2 currents:}

The recursive 
equations of motion for $h^{00}$ to order $G^2$ have already been given in Eq.~\eqref{2PMEOM}. As in Section \ref{sectionIV}, we divide up the corresponding 2PM currents according to an additional index $a$ and $b$ that tracks contributions from the mass and current multipoles separately. Then from Eq.~\eqref{2PMMassM} we get the 2PM current from iterated mass multipole pieces,
\begin{equation}\label{allordmass2PM}
\begin{split}
  J^{00[a]}_{(2)|-\boldsymbol{k}_1} 
  &=
  \frac{( 16 \pi )^2}{|\boldsymbol{k}_1|^2}
  \sum_{l,l'=0}^{\infty} \frac{(-1)^{l} i^{l+l'} M_LM_{L^{\prime}}}{l! l^\prime!} 
  \\&
  \times\int_{\boldsymbol{k_2}} \left [ -\frac{7}{8} \boldsymbol{k}_{12} {\cdot} \boldsymbol{k}_2 + \frac{5}{4} |\boldsymbol{k}_2|^2\right]   \frac{\boldsymbol{k}_{12}^L\boldsymbol{k}_{2}^{L^\prime}}{|\boldsymbol{k}_{12}|^2|\boldsymbol{k}_{2}|^2}\,,
\end{split}
\end{equation}
where $ L$, $ L' $ are multi-indices, and we write $  \boldsymbol{k}_2^e \, \boldsymbol{k}_2^{L'} =  \boldsymbol{k}_2^{e L'} $.
Considering multinomial expansion for the term $\boldsymbol{k}_{12}^{L}$,
\begin{align}
  (k_1 + k_2)^L &=\sum_{q=0}^l \binom{l}{q} k_1^{ L - Q } k_2^{Q}\,,
\label{binomial}\end{align}
where $|Q|=q$.
Working implicitly in dimensional regularization which removes all scale-free integrals, we get
\begin{equation}\label{2PM allmass}
\begin{split}
  J^{00[a]}_{(2)|-\boldsymbol{k}_1}
  =
  \frac{( 16 \pi)^2}{|\boldsymbol{k}_1|^2} 
  \!\sum_{l,l'=0}^{\infty}\sum_{q=0}^l \frac{7(-1)^{l+1} i^{l +l^\prime}}{8 \,l'!\,q!\,(l-q)!} I^{[a]}_{\boldsymbol{k}_1,q}\,,
\end{split}
\end{equation}
where 
\begin{equation}\label{nrankbubbledef}
  I^{[a]}_{\boldsymbol{k}_{1},q}
  \equiv
k_1^{eP} M_{PQ} M_{L^{\prime}}\mathcal{I}^{eL'Q}_{\boldsymbol{k}_{1}}\,,
  \quad
  L=PQ\,,
\end{equation}
and
\begin{equation}
\begin{aligned}
  \mathcal{I}^{eL'Q}_{\boldsymbol{k}_{1}}
  &=
  \int_{\boldsymbol{k}_2}\frac{k_2^{eL'Q}}{|\boldsymbol{k}_{12}|^2 \, |\boldsymbol{k}_2|^2}
  =
  \sum_{r=0}^{\lfloor\frac{n}{2}\rfloor} C_{n,r} T^{eL'Q}_{r|\boldsymbol{k}_1} |\boldsymbol{k}_{1}|^{2r-1}\,.
\end{aligned}\label{}
\end{equation}
Here $C_{n,r}$ and $T^{eL'Q}_{r|\boldsymbol{k}_1}$ are defined in \eqref{2PMBubble}. Substituting these into $I^{[a]}_{\boldsymbol{k}_{1},q}$, we have
\begin{equation}
  I^{[a]}_{\boldsymbol{k}_{1},q}
  =
  \boldsymbol{k}_1^{eP} M_{PQ} M_{L^{\prime}} \sum_{r=0}^{\lfloor\frac{n}{2}\rfloor} C_{n,r} T^{eL'Q}_{r|\boldsymbol{k}_1} |\boldsymbol{\boldsymbol{k}_1}|^{2r-1}\,.
\label{IaIntegrated}\end{equation}
The remaining task is to evaluate
\begin{equation}
  k^{eP} M_{PQ} M_{L^{\prime}} T^{eL'Q}_{r|\boldsymbol{k}}\,.
\label{}\end{equation}
At this stage the STF property is crucial. Since the mass multipoles $M_{L'}$ and $M_{L}$ are STF individually, any Kronecker delta in $T^{eL'Q}_{r|\boldsymbol{k}_1}$ that contracts two indices within $L'$, or two indices within $L=PQ$, gives zero. Hence the only non-vanishing contractions are of two types: first, Kronecker deltas connecting one index in $L'$ with one index in $Q$; second, one delta involving the distinguished index $e$, contracting $e$ with one index in $L'$ or in $Q$. 

To be specific, let $s$ denote the number of Kronecker deltas in $T^{eL'Q}_{r|\boldsymbol{k}}$ that are contracted with the mass multipoles. Type I represents the case where all $r$ Kronecker deltas are contracted with $M$, such that $r=s$. In Type II, $s$ of the Kronecker deltas are contracted with the multipole, while the single remaining delta is contracted with $M_{PQ}$ or $M_{L'}$, and $r=s+1$. Denoting the set of these contracted multi-indices by $S$, where $|S| = s$, the respective types can be written as:
\begin{equation}
\begin{aligned}
  \text{type I:} &\quad \delta^{i_{1}j_{1}} \cdots \delta^{i_{s}j_{s}} M_{i_{1}\cdots i_{s}A} M_{j_{1}\cdots j_{s}B} 
  \\&\quad
  = M_{SA} M_{SB}
  \\
  \text{type IIa:} &\quad \delta^{ee'} \delta^{i_{1}j_{1}} \cdots \delta^{i_{s}j_{s}} M_{e' i_{1}\cdots i_{s}A} M_{j_{1}\cdots j_{s}B} 
  \\&\quad
  = M_{e SA} M_{SB}
  \\
  \text{type IIb:} &\quad \delta^{ee'} \delta^{i_{1}j_{1}} \cdots \delta^{i_{s}j_{s}} M_{i_{1}\cdots i_{s}A} M_{e'j_{1}\cdots j_{s}B} 
  \\&\quad
  = M_{SA} M_{eSB}
\end{aligned}\label{}
\end{equation}
where we decomposed the multi-indices according to
\begin{equation}
\begin{aligned}
  \text{type I:} &\quad
  L^{\prime}=S A, \quad Q=S R, 
  \\&\quad
  L=P Q=S B, \quad B=PR\,,
  \\
  \text{type IIa:} &\quad
  L^{\prime}=eS A, \quad Q=S R,
  \\&\quad
  L=P Q=S B, \quad B=PR\,,
  \\
  \text{type IIb:} &\quad
  L^{\prime}=S A, \quad Q=e S R,
  \\&\quad
  L=P Q=eS B, \quad B=PR\,,
\end{aligned}\label{}
\end{equation}
Contracting with all the remaining $k$ vectors, every surviving term reduces to the common tensor structure $M_{S A} k^{A} M_{S B} k^{B}$ by symmetry. 

We now proceed to evaluate the contractions explicitly, first considering type I. In this case all $s$ deltas in $T^{N}_{r|k}$ must connect $L'$ with $Q$. After contraction with the external factors $k^{e}k^{P}$, one obtains
\begin{equation}
\begin{aligned}
  &k^{eP} T_{s|k}^{e L'Q} M_{L^{\prime}} M_{PQ}
  =
  N_{1}|k|^{2} M_{S A} k^{A} M_{S B} k^{B}\,,
  \\
  &N_{1}
  = \binom{l^{\prime}}{s}\binom{q}{s} s!
  = \frac{l^{\prime}!q!}{s!\left(l^{\prime}-s\right)!(q-s)!}\,.
\end{aligned}\label{IaPart1}
\end{equation}
We next consider type II. Besides the $s$ contractions between $L'$ and $Q$, there must be one additional Kronecket delta involving the index $e$. For type IIa case, the number of such terms is 
\begin{equation}
\begin{aligned}
  N_{2}
  &=
  \frac{l'!q!}{s!\left(l'-s-1\right)!(q-s)!}=\left(l^{\prime}-s\right) N_{1}\,.
\end{aligned}\label{}
\end{equation}
For type IIb case, the number of those terms is 
\begin{equation}
\begin{aligned}
  N_{2'}
  &=
  \frac{l'!q!}{s!\left(l'-s\right)!(q-s-1)!}=(q-s) N_{1}\,.
\end{aligned}\label{}
\end{equation}
Combining these two results, we have
\begin{equation}
\begin{aligned}
  &k^{a} k^{P} T_{s+1|k}^{a L^{\prime} Q} M_{L^{\prime}} M_{PQ}
  \\
  &=
  \big[N_{2}+N_{2'}\big] M_{S A} k^{A} M_{S B} k^{B}\,,
  \\
  &=
  (l^{\prime}+q-2 s) N_{1} M_{S A} k^{A} M_{S B} k^{B}\,.
\end{aligned}\label{IaPart2}
\end{equation}
Finally, combining \eqref{IaPart1} and \eqref{IaPart2}, we find
\begin{equation}
\begin{aligned}
  I^{[a]}_{\boldsymbol{k},q}
  &=
  (-1)^{n}\!\! \sum_{s=0}^{\min (l'\!, q)}\!\! N_{1}
  \left[C_{n,s}+(l'+q-2 s) C_{n,s+1}\right] 
  \\&\qquad\qquad\qquad
  \times|\boldsymbol{k}|^{2 s+1} M_{S A} k^{A} M_{S B} k^{B}\,.
\end{aligned}\label{}
\end{equation}
Evaluating $C_{n,s}+(l'+q-2 s) C_{n,s+1}$, we obtain
\begin{equation}
  C_{n,s}+(l'+q-2 s) C_{n,s+1}
  =
  \frac{(-1)^{s} \Gamma\left(l'{+}q{-}s+\frac{1}{2}\right)}{2^{s+4} \sqrt{\pi} \Gamma(l'{+}q{+}1)}\,.
\label{}\end{equation}
It follows that $I^{[a]}_{\boldsymbol{k},q}$ is given by
\begin{equation}
\begin{aligned}
  I^{[a]}_{\boldsymbol{k},q}
  &=
  \sum_{s=0}^{\min (l^{\prime}, q)}
  C^{[a]}_{n,s}\, |\boldsymbol{k}|^{2 s+1} M_{SA} k^{A} M_{SB} k^{B}\,,
  \\
  C^{[a]}_{n,s}
  &=
  \frac{(-1)^{n+s} l^{\prime}!q!}{2^{s+4} \sqrt{\pi} s! (l^{\prime}-s)!(q-s)!} 
  \frac{\Gamma\left(n-s-\frac{1}{2}\right)}{\Gamma(n)}\,,
\end{aligned}\label{IntegralOfJ00a}
\end{equation}
where $n=l^{\prime}+q+1$. Substituting the results into \eqref{2PM allmass}, we obtain
\begin{equation}\label{2PM_all_a}
\begin{split}
  J^{00[a]}_{(2)|-\boldsymbol{k}_1}
  &=
  \frac{( 16 \pi)^2}{|\boldsymbol{k}_1|^2} \!\!\sum_{l,l'=0}^{\infty} \frac{7(-1)^{l+1} i^{l +l^\prime}}{8 \; l! l^\prime!}\!\!
  \sum_{q=0}^l \binom{l}{q} 
  \\&\quad
  \times\sum_{s=0}^{\min (l^{\prime}, q)}
  C^{[a]}_{n,s}\, |\boldsymbol{k}_1|^{2 s+1} M_{SA} k^{A}_1 M_{SB} k^{B}_1 \,.
\end{split}
\end{equation}

Eq.~\eqref{2PM allmass} provides us with the full 2PM contribution from all mass multipoles in terms of a generalized (tensor) bubble integral. We now turn to those coming from current multipoles. We read off the
corresponding 2PM pieces for $h_{(2)}^{00}$ from Eq.~\eqref{2PMEOM}, giving us the corresponding momentum space integral
\begin{equation}\label{a2PMcmultipole}
\begin{split}
  J^{00[b]}_{(2)|-\boldsymbol{k}_1}
  &{=}
  \frac{1}{|\boldsymbol{k}_1|^2} \int_{\boldsymbol{k_2}} \bigg[
    \frac{\boldsymbol{k}_{1} {\cdot} \boldsymbol{k}_2 - |\boldsymbol{k}_2|^2}{2} J^{0i}_{(1)|-\boldsymbol{k}_{12}} J^{0i}_{(1)|\boldsymbol{k}_2}
  \\&\qquad\qquad\quad
  + \frac{3}{2} \boldsymbol{k}_{2}^i\boldsymbol{k}_{12}^j J^{0i}_{(1)|-\boldsymbol{k}_{12}} J^{0j}_{(1)|\boldsymbol{k}_2}\bigg]\,.
\end{split}
\end{equation}
As initial conditions, we have the $1$PM currents,
\begin{equation}
  J^{0i}_{(1)|\pm\boldsymbol{k}}
  =  16\pi\sum_{l = 1}^{\infty} l (\pm i)^{l}\frac{[\tilde{\boldsymbol{S}}\times \boldsymbol{k}]^{L-1}_{i}}{(l+1)! |\boldsymbol{k}|^2}\,,
\end{equation}
where 
\begin{equation}
  \big[\tilde{\boldsymbol{S}}\times \boldsymbol{k}\big]^{L-1}_{i} \equiv \epsilon_{iab}S_{bL-1} k^{aL-1}\,.
\label{}\end{equation}
Using the derivative acting on the exponentials and removing the scaleless integral,
\begin{equation}
\begin{aligned}
  &J^{00[b]}_{(2)|-\boldsymbol{k}_1} 
  =
  \frac{(16 \pi)^2}{|\boldsymbol{k}_1|^2} \sum_{l,l' = 1 }^{\infty} 
  \frac{(-1)^{l} i^{l+l'}ll'}{(l+1)!(l'+1)!} I^{[b]}_{\boldsymbol{k}_{1}}\,,
\end{aligned}\label{step1incurrent}
\end{equation}
where
\begin{equation}
  I^{[b]}_{\boldsymbol{k_{1}}}
  {=}
  \int_{\boldsymbol{k_2}}\!
  \frac{\big(\boldsymbol{k}_{1} {\cdot} \boldsymbol{k}_2 \delta^{ij} {+} 3k_{2}^{i} k_{12}^{j} \big) [\tilde{\boldsymbol{S}}{\times}\boldsymbol{k}_{12}]^{L{-}1}_{i}[\tilde{\boldsymbol{S}}{\times}\boldsymbol{k}_{2}]^{L'\!-1}_{j}}{2|\boldsymbol{k}_{12}|^2|\boldsymbol{k}_2|^2}\,.
\label{}\end{equation}
Using the following identity,
\begin{equation}
\begin{aligned}
  k_{2}^{i}[\tilde{\boldsymbol{S}} \times\boldsymbol{k}_{12})]_{i}^{L-1}
  &=
  -k_{1}^{i}[\tilde{\boldsymbol{S}} \times\boldsymbol{k}_{12}]_{i}^{L-1}\,,
  \\
  k_{12}^{j}[\tilde{\boldsymbol{S}} \times \boldsymbol{k}_{2}]_{j}^{L'-1}
  &=
  k_{1}^{j}[\tilde{\boldsymbol{S}} \times \boldsymbol{k}_{2}]_{j}^{L'-1}\,,
\end{aligned}\label{}
\end{equation}
we can reduce the integral into
\begin{equation}
  I^{[b]}_{\boldsymbol{k}_{1}}
  =
  \!\int_{\boldsymbol{k}_{2}}\! \frac{\big(\boldsymbol{k}_{1} {\cdot} \boldsymbol{k}_{2} \delta^{i j} {-}3 k_{1}^{i} k_{1}^{j}\big)[\tilde{\boldsymbol{S}} {\times}\boldsymbol{k}_{12}]_{i}^{L-1}[\tilde{\boldsymbol{S}} {\times} \boldsymbol{k}_{2}]_{j}^{L'-1}}{2|\boldsymbol{k}_{2}|^{2}|\boldsymbol{k}_{12}|^{2}}\,.
\label{}\end{equation}
Using the binomial expansion \eqref{binomial}, we have
\begin{equation}
\begin{aligned}
  k_{12}^{a(L-1)}
  =
  \sum_{p=0}^{l-1}\binom{l{-}1}{p}\left(k_{1}^{aP} k_{2}^{Q} +k_{1}^{P} k_{2}^{a Q}\,.\right)
\end{aligned}
\end{equation}
where $L-1 = PQ $, $|P| = p$ and $|Q| = l-1-p$. Now, $I^{[b]}_{\boldsymbol{k}}$ can be decomposed into two parts:
\begin{equation}
\begin{aligned}
  I^{[b]}_{\boldsymbol{k}}
  &=
  \frac{1}{2}\sum_{p=0}^{l-1}\binom{l{-}1}{p} 
  \Big[\,
      f_{(1)}^{aN}\mathcal{I}_{\boldsymbol{k}}^{a N}
    + f_{(2)}^{abN}\mathcal{I}_{\boldsymbol{k}}^{a b N}\,
  \Big]\,,
\end{aligned}\label{}
\end{equation}
where $\mathcal{I}^{N}_{\boldsymbol{k}}$ is defined in \eqref{2PMBubble} and
\begin{equation}
\begin{aligned}
  f_{(1)}^{aN}
  &=
  (-1)^{n+1}\Big(k^{(a|}k_{b} Y^{b|N)}_{ii} -3k^{i} k^{j} Y^{(aN)}_{ij}\Big)\,,
  \\
  f_{(2)}^{abN}
  &=
  (-1)^{n+2} k^{(a} Y^{b N)}_{ii} \,,
\end{aligned}\label{f_tensors}
\end{equation}
and $Y^{a N}_{ij}$ is a multiplication of two spin multipoles
\begin{equation}
\begin{aligned}
  Y^{aN}_{ij}\equiv Y^{a Q c (L'-1)}_{ij} &= k^{P}\epsilon_{i ab} \epsilon_{jcd} S_{bPQ} S_{d (L'-1)}\,.
\end{aligned}\label{}
\end{equation}
Note that the parentheses on the right-hand side in \eqref{f_tensors} imply the total symmetrization of $a$ and all the multi-indices of $N$ for the first line, and total symmetrization of $a$, $b$ and all the multi-indices of $N$ for the second line. 
Thus we now want evaluate $f_{(1)}^{aN}$ and $f_{(2)}^{abN}$ by eliminating $T^{N}_{r}$. Let us consider an arbitrary symmetric tensor $X_{N}$ and its contraction with the $T^{N}_{r|\boldsymbol{k}}$
\begin{equation}
  X_{N} T^{N}_{r|\boldsymbol{k}}
  =
  \frac{n!}{ (2r)!(n-2 r)!}\left(\operatorname{Tr}^{r} X\right)_{Q} k^{Q}\,,
\label{}\end{equation}
where $N= i_{1} i_{2} \cdots i_{2r} Q$ and
\begin{equation}
  \left(\operatorname{Tr}^{r} X\right)_{Q} \equiv X_{a_{1} a_{1} \cdots a_{r} a_{r} Q}\,.
\label{}\end{equation}
is an $r$-fold reduced trace in the form shown. The full contraction between $X_{N}$ and $\mathcal{I}_{\boldsymbol{k}}^{N}$ is thus given by
\begin{equation}
\begin{aligned}
  X_{N} \mathcal{I}^{N}_{\boldsymbol{k}}
  &=
  \sum_{r=0}^{\lfloor n / 2\rfloor} C^{[b]}_{n, r} |\boldsymbol{k}|^{2 r-1}\left(\operatorname{Tr}^{r} X\right)_{Q} k^{Q}\,,
\end{aligned}\label{}
\end{equation}
where 
\begin{equation}
  C^{[b]}_{n,r}
  =
  (-1)^{r} \frac{\Gamma\left(n-r+\frac{1}{2}\right)}{2^{r+3} \sqrt{\pi} (2r)!(n-2 r)!}\,.
\label{}\end{equation}

Then $I^{[b]}_{\boldsymbol{k}}$ becomes
\begin{equation}
\begin{aligned}
  I^{[b]}_{\boldsymbol{k}}
  &=
  \frac{1}{2} \sum_{p=0}^{l-1}\binom{l{-}1}{p}
  \Bigg[
    \sum_{r=0}^{\lfloor \frac{n+1}{2}\rfloor} \frac{C^{[b]}_{n+1, r}}{|\boldsymbol{k}|^{1-2 r}}  \big({\rm Tr}^{r} \!f_{(1)}\big)_{Q} k^{Q}
  \\&\qquad\qquad
    + \sum_{r'=0}^{\lfloor \frac{n}{2}{+}1\rfloor} \frac{C^{[b]}_{n+2, r'}}{|\boldsymbol{k}|^{1-2 r'}} \big({\rm Tr}^{r'}\!f_{(2)}\big)_{Q'} k^{Q'}
  \Bigg]\,.
\end{aligned}\label{}
\end{equation}

Hence, by combining terms, we have total $J^{00}_{(2)|-\boldsymbol{k}}$ as the sum of the two contributions,
\begin{widetext}
\begin{equation}
\begin{aligned}
  J^{00}_{(2)|-\boldsymbol{k}}
  &=
  J^{00[a]}_{(2)|-\boldsymbol{k}}+J^{00[b]}_{(2)|-\boldsymbol{k}}
  \\
  &=
  \frac{( 16 \pi)^2}{|\boldsymbol{k}|^2} \!\!\sum_{l,l'=0}^{\infty} \frac{7(-1)^{l+1} i^{l +l^\prime}}{8 \; l! l^\prime!}\!\!
  \sum_{q=0}^l \binom{l}{q} \sum_{s=0}^{\min (l^{\prime}, q)}
  C^{[a]}_{n,s}\, |\boldsymbol{k}|^{2 s+1} M_{SA} k^{A} M_{SB} k^{B}
  \\&
  +
  \frac{(16 \pi)^2}{2|\boldsymbol{k}|^2} \sum_{l,l' = 1 }^{\infty} 
  \frac{(-1)^{l} i^{l+l'}ll'}{(l+1)!(l'+1)!} 
  \sum_{p=0}^{l-1}\binom{l{-}1}{p}
  \Bigg[
    \sum_{r=0}^{\lfloor \frac{n+1}{2}\rfloor} \frac{C^{[b]}_{n+1, r}}{|\boldsymbol{k}|^{1-2 r}}  \big({\rm Tr}^{r} \!f_{(1)}\big)_{A} k^{A}
   + \sum_{r'=0}^{\lfloor \frac{n}{2}{+}1\rfloor} \frac{C^{[b]}_{n+2, r'}}{|\boldsymbol{k}|^{1-2 r'}} \big({\rm Tr}^{r'}\!f_{(2)}\big)_{B} k^{B}
  \Bigg]\,,
\end{aligned}\label{}
\end{equation}
\end{widetext}
This rather formidable expression gives the current $J^{00}_{(2)|-\boldsymbol{k}}$ at second post-Minkowskian order and including all mass and current multipoles.
We have checked this general formula by truncating the expansion at to $l = l' = 2$ for $J^{[a]}_{(2)|-\boldsymbol{k}}$ and $l = l' = 1$ for 
$J^{[b]}_{(2)|-\boldsymbol{k}}$. This reproduces correctly the result of Eq. \eqref{Joi2PM}. 

Having obtained the full current $J^{00}_{(2)|-\boldsymbol{k}}$ at 2PM order, the metric components$h^{00}_{(2)}$ can be straightforwardly obtained by extracting the powers of $|\boldsymbol{k}|$ and $k$-components from the above results, collecting them, and applying the Fourier transform formula in \eqref{FourierFormula}.
%

\section{Conclusion}

Extending the recursive method of ref. \cite{Damgaard:2024fqj}, we have set up the iterative equations that allow us to compute the gravitational metric of spatially localized and stationary sources to high order in $G$. For a point-like source, the method allows for a computation to all orders in the coupling $G$, for which the Schwarzschild metric is recovered for the given gauge in the full domain of convergence of the perturbatively expanded solution. For a black hole with spin, which is sourced by a more complicated energy-momentum tensor, the iterative solution becomes more cumbersome but it can again be arranged to essentially yield the metric to any desired orders in $G$ and spin $a$. Applying the same approach to a more general spatially localized matter source, as done in this paper, shows the versatility of such an iterative scheme that is based entirely on the classical equations of motion without reference to the effective field theory of point-like massive sources.

The general multipole expansion can be expressed in terms of an infinite series of mass multipoles and current multipoles. In the Landau-Lifshitz 
formalism and with the de Donder gauge condition that we impose on $h^{\mu\nu}$, the leading-order multipole expansion yields order-$G$ 
contributions to $h^{00}$ and $h^{0i}$ but none for $h^{ij}$. The latter only builds up at order $G^2$ (and recursively at higher orders) through 
combinations of the lower-order terms from $h^{00}$ and $h^{0i}$. In a highly degenerate corner of the space of multipoles, we recover the Kerr metric
in harmonic coordinates to the given order. Interestingly, the gauge ambiguity that remains even in the chosen harmonic gauge \cite{Fromholz:2013hka}
shows up already in the current-current contributions to $h^{ij}$ at order $G^2$ where our iterative scheme reproduces the ``exact'' Kerr metric
only up to terms that can be removed by a coordinate shift consistent with the chosen gauge.

After displaying the general iterative equations from which we can push the general multipole computation to as high orders we wish, we have 
truncated explicit 2PM results at mass quadrupole order and leading current multipole order simply because higher-order terms become analytically quite unwieldy on account of the numerous different tensor
structures that arise in the chosen coordinates. Although the subject of our paper obviously is far from new, we have tried to convey that our computational
methods are. By Fourier transforming from metrics to currents the iterative structure that is built into the Einstein equation in perturbation theory become much simplified, and all
involved integrals become analytically doable. This is where we can capitalize on all the recent developments in post-Minkowskian perturbation theory based on amplitude or worldline methods. In fact, for the multipole expansion considered here the required integrals in momentum space (loops in the language of quantum field theory) remain exceedingly
simple in comparison with the integrals encountered in the two-body problem. The origin of this simplicity is easily identified: all pertinent integrals correspond to what in quantum
field theory would be considered tensor integrals of massless propagators, $i.e.$ generalized tensor bubble integrals.  An interesting question that goes beyond computation of the gravitational metric
of the compact object is how tidal deformations (and hence also Love numbers) can be linked to the multipole expansion, thus providing physical and measurable
consequences of the expansion. Looking further ahead would be the
extension of the iterative setup to the non-stationary case with, in particular, applications to the bound two-body problem in general relativity.

\begin{acknowledgements}
One of us (PHD) would like to thank Luis Lehner and Suvendu Giri for discussions and for hospitality at Perimeter Institute. TR gratefully acknowledges the hospitality of NBIA, Copenhagen, during the early stages of this work, as well as support from APCTP, Pohang, where part of this work was conducted. TR also thanks Raj Patil, Kwangeon Kim, and Thibault Damour for insightful discussions. This work was supported by the National Research Foundation of Korea(NRF) grant funded by the Korea government(MSIT) RS-2025-24533223.  HL is supported by KIAS Individual Grant PG105701. KL is supported by KIAS Individual Grant QP106001 and TR is supported by KIAS Individual Grant QP108201 via the Quantum Universe Center at KIAS. 
 
\end{acknowledgements}

\appendix
\section{Useful STF tensor identities}\label{appendixA}

In this appendix, we collect some useful relations among STF tensors.
For a general rank-$p$ tensor $ A^{i_1 \cdots i_p} $, the STF projection can be written as \cite{Thorne:1980ru,Blanchet:1985sp,Damour:1990gj}:
\begin{equation}\label{STFdef}
 \hat{A}^{ i_1 \cdots i_p} 
= \sum_{k=0}^{\left\lfloor \frac{p}{2} \right\rfloor} \alpha_{p,k}\ 
\delta^{(i_1 i_2} \cdots \delta^{i_{2k-1} i_{2k}} \, 
S^{i_{2k+1} \cdots i_p)j_1 j_1 \cdots j_k j_k}
\end{equation}
 where $S^{i_1 \cdots i_p} = A^{(i_1 \cdots i_p)}$ is the fully symmetrized $A$-tensor. The $k$'th term in the sum has $k$ pairs of indices contracted with 
 kronecker-deltas as indicated, and the coefficients $\alpha_k$ are given by
 $$
\alpha_{p,k}= \frac{(-1)^k \, p! \, (2p - 2k - 1)!!}{(2p - 1)!! \, (2k)!! \, (p - 2k)!} ~.
$$

\paragraph*{$k$-fold trace:}
For a fully symmetric rank-$p$ tensor $S^{i_1 \dots i_p}$, one can define a
$k$-fold trace ($\mathrm{Tr}^k$) by contracting $k$ independent 
pairs of indices:
\begin{equation}
(\mathrm{Tr}^k S)^{\, i_{2k+1}\dots i_p}
\equiv
\delta_{i_1 i_2}\,
\cdots\,
\delta_{i_{2k-1} i_{2k}}\,
S^{i_1 \dots i_{2k} \, i_{2k+1}\dots i_p}.
\end{equation}

Each trace contracts two indices, so a $k$-fold trace removes $2k$ indices 
and reduces the rank from $p$ to $p-2k$, 
with $ 0 \le k \le \left\lfloor \frac{p}{2} \right\rfloor$
.
\subsection*{Explicit STF projections up to rank 4}\label{}
\paragraph{Rank 1 ($p=1$)}: \begin{equation}
    \hat{A}^{ i }
= S^{i}
= A^{i}\, ,
\end{equation}
\paragraph{Rank 2 ($p=2$)}: 
\begin{equation}
    \hat{A}^{ ij }
=
S^{ij}
-\frac{1}{3}\,\delta^{ij}\,S^{aa}\, ,
\end{equation}
\paragraph{Rank 3 ($p=3$)}:
\begin{equation}
\hat{A}^{ ijk }
=
S^{ijk}
-\frac{1}{5}\Big(
\delta^{ij} S^{aa k}
+\delta^{ik} S^{aa j}
+\delta^{jk} S^{aa i}
\Big)\,,
\end{equation}
\paragraph{Rank 4 ($p=4$)}
\begin{equation}\label{rank4stf}
    \begin{split}
        \hat{A}^{ ijkl}
&=
S^{ijkl}
-\frac{1}{7}\Big(
\delta^{ij} S^{aa kl}
+\delta^{ik} S^{aa jl}
+\delta^{il} S^{aa jk}\\
&
+\delta^{jk} S^{aa il}
+\delta^{jl} S^{aa ik}
+\delta^{kl} S^{aa ij}
\Big)\\
&
+\frac{1}{35}\Big(
\delta^{ij}\delta^{kl}
+\delta^{ik}\delta^{jl}
+\delta^{il}\delta^{jk}
\Big)\,S^{aabb}.
\end{split}
\end{equation}
with repeated indices summed over.

\subsection*{Scalar Products and STF Tensors}

The two alternative descriptions of multipoles in terms of either spherical harmonics or STF tensors are a change of basis, 
and we can go back and forth between them. This is all clearly explained in Ref. \cite{Thorne:1980ru} and we shall just summarize a few main points here.
Let
\begin{equation*}
  \mathcal H_l
  \equiv  \{\, A^{i_1\cdots i_l}\ \text{symmetric and trace-free in } \mathbb R^3 \,\}.
\end{equation*}
be a set of rank-$\ell$ STF tensors.
This space carries an irreducible representation of $SO(3)$ of dimension
\begin{equation*}
    \dim \mathcal H_l = 2l+1,
\end{equation*}
equivalent to the usual spin-$\ell$ representation spanned by the spherical harmonics
$\{Y_{l m}\}_{m=-l}^{l}$\,.

According to the Legendre polynomial addition theorem, $P_{l}(\cos \theta)$ can be written in terms of spherical harmonics
\begin{equation}\label{Lpol}
  P_l(\cos\theta)
  =
  \frac{4 \pi}{2l + 1} \sum_{m= -l}^{l} Y_{l m} (\theta_1,\phi_1)Y_{l m}^{*} (\theta_2, \phi_2 )
\end{equation}
where $\cos\theta = \cos\theta_1\cos\theta_2 + \sin\theta_1\sin\theta_2\cos(\phi_1-\phi_2)$.
These spherical harmonics can be written in terms of the STF tensor basis as follows \cite{Thorne:1980ru, Hartmann1994} : 
\begin{equation}\label{nhat}  Y_{l m} (\theta_1,\phi_1)
  =
  Y_{l m}^{L} \; \hat{n}^{L}(\theta_1,\phi_1)
\end{equation}
where the coefficients $ Y_{l m}^{L}$ are independent of the coordinates of the sphere, and $L$ is again a multi-index so that $Y_{l m}^{L} =Y_{l m}^{i_1, \cdots i_{L}} $. The Cartesian tensors, $\hat{n}^{L}$ are irreducible representations of the tensors, and very useful for the analysis. An important property of these tensors is that they generate harmonic functions. Indeed, the solid harmonics $ r^{l} \hat{n}^{L}$, satisfy the flat-space Laplace equation,
\begin{equation}
    \nabla^2 \big( r^{l} \hat{n}^{L} \big) = 0,
\end{equation}
which makes them the natural cartesian analogue of the spherical harmonics. The scalar harmonics are orthogonal and satisfy the identity,
\begin{equation}
  Y^{l m *}_{L}\,Y^{l m'}_{L}
  =
  \frac{(2\ell+1)!!}{4\pi\,l!}\,\delta_{m m'}\,.
\end{equation}
Using this the above relation in Eq.~\eqref{nhat}, can be written as, 
\begin{equation}
\begin{aligned}
  & \sum_{m= -l}^{l} Y_{l m} (\theta_1,\phi_1)Y_{l m}^{*} (\theta_2, \phi_2 )  \\
  &=
  \hat{n}^{L}(\theta_1,\phi_1) \sum_{m= -l}^{l} Y_{l m}^{L}Y_{l m}^{*} (\theta_2, \phi_2 ) 
  \\
  &=\frac{(2l +1)!!}{4 \pi l !} \hat{n}^{L}(\theta_1,\phi_1) \hat{n}^{\prime L}(\theta_2,\phi_2) 
\end{aligned}\label{}
\end{equation}
Hence, \eqref{Lpol} can be written as
\begin{align}\label{poly}
  P_l(\cos\theta)
  =
  \frac{(2l -1)!!}{l !} \hat{n}^{L}(\theta_1,\phi_1) \hat{n}^{\prime L}(\theta_2,\phi_2). 
\end{align}
These identities can be used to replace angular functions in multipole expansions with STF tensor structures, and they are essential when converting expressions into the STF basis.

\section{Fourier Transforms and STF Tensors}\label{FTDtrick}

The STF tensor formalism beautifully translates into momentum space through Fourier transforms. This is essential for our computations below, and we shall therefore
spell it out in some detail. To start, consider the first relevant Fourier transform, corresponding to the monopole term $1/r$,
\begin{equation}
   I
   \equiv
   \int \mathrm{d}^3 \boldsymbol{x} \, \frac{e^{-i \boldsymbol{k} \cdot \boldsymbol{x}}}{r} 
   =
   \frac{4\pi}{|\boldsymbol{k}|^{2}} \,.
\label{}\end{equation}
All Fourier transforms are defined consistently as
\begin{equation}
  \tilde f(\boldsymbol{k})
  =
  \int_{\boldsymbol{x}} e^{-i \boldsymbol{k} \cdot \boldsymbol{x}} f(\boldsymbol{x}), \qquad
  f(\boldsymbol{x})
  =
  \int_{\boldsymbol{k}} e^{i \boldsymbol{k} \cdot \boldsymbol{x}} \tilde f(\boldsymbol{k}),
\end{equation}
where
\begin{equation}
  \int_{\boldsymbol{x}} \equiv \int d^3 \boldsymbol{x}
  \qquad
  \int_{\boldsymbol{k}}\equiv \int \frac{d^3 \boldsymbol{k}}{(2\pi)^3}
\end{equation}
so that standard Fourier pairs such as
\begin{equation}
  \int d^3 \boldsymbol{x}
  \frac{e^{-i \boldsymbol{k} \cdot \boldsymbol{x}}}{r}
  = \frac{4 \pi}{|\boldsymbol{k}|^2}, 
  \qquad
  \int \frac{d^3 \boldsymbol{k}}{(2\pi)^3} \, \frac{e^{i \boldsymbol{k} \cdot \boldsymbol{x}}}{|\boldsymbol{k}|^2} = \frac{1}{4 \pi r}
\end{equation}
are consistent
The integral extends over all of three-space, including the origin at $r=0$. 
Intuitively, when we seek solutions to the Einstein equations far away from the matter
sources with the center of mass at the origin of coordinates, we would not expect to be concerned with the point $r=0$. Yet, by first Fourier transforming (and subsequently
taking the inverse Fourier transform back to real space), we do encounter this spurious point. We therefore need to handle it with care and also provide expressions
for $\partial_L(1/r)$, where $\partial_{L} = \partial_{i_{1}} \cdots \partial_{i_{l}}$ for $L = i_{1} \cdots i_{l}$, that are not restricted to $r>0$. As a starting point, consider
\begin{align}
  \partial_a \partial_b \frac{1}{r}
  &=
  - \frac{4 \pi}{3} \delta^{ab} \delta(\boldsymbol{x}) 
  + \frac{1}{r^3}\left( \frac{3 x^a x^b}{r^2}  - \delta^{ab}\right) ~,
\end{align}
where the first term is required for consistency with the point-source Poisson equation 
\begin{equation}
\partial^2 \frac{1}{r} = -4\pi\delta (\boldsymbol{x}) ~.
\end{equation}
Hence, in terms of the Fourier integral 
\begin{equation}
  \partial_{ab} \frac{1}{r}
  =  
  \partial_{a b} \int_{\boldsymbol{k}} \, e^{i \boldsymbol{k} \cdot \boldsymbol{x}}\frac{4 \pi}{|\boldsymbol{k}|^2}
  =
  - \frac{4 \pi}{3} \delta^{ab} \delta(\boldsymbol{x})
  + \frac{3}{r^3} \hat{n}^{ab},
\end{equation}
which implies
\beq
 \int_{\boldsymbol{k}} \, e^{i \boldsymbol{k} \cdot \boldsymbol{x}}4 \pi \left( \frac{ k^a k^b}{|\boldsymbol{k}|^2} - \frac{1}{3}  \delta^{ab}\right)
     = - \frac{3}{r^3} \hat{n}^{ab}
\eeq
and hence 
\beq
   \int_{\boldsymbol{k}} \, e^{i \boldsymbol{k} \cdot \boldsymbol{x}} \frac{ 4 \pi\hat{k}^{ab}}{|\boldsymbol{k}|^2} 
     = - \frac{3}{r^3} \hat{n}^{ab} 
\label{kkintegral}\eeq
where
\begin{equation}
\begin{aligned}
  \hat{n}^{ab} &\equiv \frac{x^{\langle a}x^{b\rangle}}{r^2}  = \frac{1}{r^2} \left( x^a x^b  - \frac{1}{3} r^2 \delta^{ab}\right)\,,
  \\
  \hat{k}^{ab} &\equiv k^{a} k^{b}- \frac{1}{3}|\boldsymbol{k}|^2 \delta^{ab} ~.
\end{aligned}\label{}
\end{equation}
This can be generalized to Fourier transforms with a multi-index $L$-derivative
\begin{equation}
\int_{\boldsymbol{k}} \, \frac{\partial_{L} \left(e^{i \boldsymbol{k} \cdot \boldsymbol{x}}\right)}{|\boldsymbol{k}|^2}   
  = \partial_{L}\left( \frac{1}{4\pi r}\right) 
\label{Fourier_transform_identity}\end{equation}
which in turn can be used to compute Fourier transforms of tensor integrals. We shall return to this below.
We note that we can absorb the local $\delta$-function terms by redefining the derivative tensors on the 
left-hand side by subtracting the contribution at $r=0$,
\begin{equation}\label{derac1}
\begin{split}
  \hat{\partial}_{ab} \frac{1}{r}&=    \frac{1}{r^3}\left( \frac{3 x^a x^b}{r^2}  - \delta^{ab}\right)= \frac{3}{r^3} \hat{n}^{ab}
    \end{split}
\end{equation}
where $\hat{\partial}_{ab}$ can be interpreted as a definition of the second rank STF derivative operators
\begin{equation}\label{derivativeop}
  \hat{\partial}_{ab} \equiv \partial_a \partial_b - \frac{1}{3} \delta^{ab}\partial^2\,.
\end{equation}
In particular, this implies,
\begin{equation}
  \int_{\boldsymbol{x} }\, e^{-i \boldsymbol{k} \cdot \boldsymbol{x}} \frac{1}{r^3} \hat{n}^{ab}   = 
-\frac{ 4 \pi}{3|\boldsymbol{k}|^2} \hat{k}^{ab}\,,
\end{equation}
so that the STF-property is being preserved under Fourier transformations, as it should. 

In general, the Fourier transform $\mathcal{F}$ in $d$ dimensions commutes with $SO(d)$ rotations, i.e., $\mathcal{R}[\mathcal{F}[f]] = \mathcal{F}[\mathcal{R}[f]]$ for any $\mathcal{R} \in SO(d)$, and therefore preserves the irreducible representations of $SO(d)$. Since STF tensors form irreducible representations of $SO(d)$, the Fourier transform of any STF tensor must itself remain an STF tensor. This fact greatly constrains the structure of the result: the $d$-dimensional Fourier transform of $\hat{x}^L/|\boldsymbol{x}|^n$, where $\boldsymbol{x}$ is a $d$-dimensional position vector, must take the form $\hat{k}^L \cdot f(|\boldsymbol{k}|)$ for some scalar function $f$. The explicit form of $f$ can be determined by employing the Schwinger parametrization together with the $d$-dimensional Gaussian integral formula as
\begin{equation}
  \int d^{d} \boldsymbol{x} e^{-i \boldsymbol{k} \cdot \boldsymbol{x}} 
  \frac{\hat{x}^{L}}{r^{m}}
  = 
  \frac{\pi^{\frac{d}{2}}\Gamma\left(\frac{d-m+2l}{2}\right)}{i^{l}2^{m-d-l}\Gamma\left(\frac{m}{2}\right)} \frac{\hat{k}^L}{|\boldsymbol{k}|^{d-m+2l}}\,.
\label{STFFourierTransf}\end{equation}
The corresponding $d$-dimensional Fourier transform of an arbitrary tensor field without the STF projection is given in Appendix \ref{FourierFormulaPart}.


%
For stationary sources, the Einstein equations of motion at the linearized level reduce to
\begin{align}
  \nabla^2 h^{\mu \nu}_{(1)}(\boldsymbol{x}) = -16 \pi T^{\mu \nu}(\boldsymbol{x}) \,, \label{h1eq}
\end{align}
which in Cartesian coordinates is solved by
\begin{equation}
  h^{\mu \nu}_{(1)}(\boldsymbol{x})
  =
  4\int d^3 \boldsymbol{x}'\, \frac{T^{\mu \nu}(\boldsymbol{x}')}{\lvert \boldsymbol{x} - \boldsymbol{x}' \rvert},
\label{solPoisson}\end{equation}
Here, we assume that the source is spatially confined to a bounded region $\lvert \boldsymbol{x}'\rvert < r_0 $ for some given $r_0$ so $T^{\mu \nu}(\boldsymbol{x}')=0$ for $\lvert \boldsymbol{x}'\rvert>r_0$. Exterior to the source the solution thus admits a standard multipole expansion for $r = |\boldsymbol{x}| > |\boldsymbol{x}'|$ based on Taylor-expanding:
\begin{equation}
   \frac{1}{|\boldsymbol{x}-\boldsymbol{x}'|}
   =
   \sum_{l=0}^{\infty}\frac{(-1)^{l}}{l!}\,\boldsymbol{x}'^{L}\,\partial_{L}\!\left(\frac{1}{r}\right).
\label{taylorExpansion}\end{equation}
The expansion simplifies considerably since by assumption all multipole moments are time-independent.
To leading order, ignoring gravitational self-interactions and, using the definition of multipoles given in section \ref{defsec} the solution
takes the form of (using a notation in which multi-index $aL-1 = ai_2\dots i_{l}$):
\begin{equation}
\begin{aligned}
  h_{(1)}^{00}(\boldsymbol{x})
  &=
  4\sum_{l=0}^{\infty} \frac{(-1)^l}{l!} \, \partial_L \, \frac{M_L}{r}\,,
  \\
  h_{(1)}^{0i}(\boldsymbol{x})
  &= 
  4 \sum_{l=1}^{\infty} \frac{l(-1)^l}{(l+1)!} \, \epsilon_{iab} \, \partial_{aL-1} \, \frac{S_{bL-1}}{r}\,,
  \\
  h_{(1)}^{ij}(\boldsymbol{x})
  &= 0\,. 
\end{aligned}\label{h1sol}
\end{equation}
in what is known as the canonical basis. Note that the contraction between $x'^{L}$ and $\partial_{L}(1/r)$ in Eq.~\eqref{taylorExpansion} satisfies
\begin{equation}
  x'^{L} \partial_{L}\frac{1}{r} 
  =
  \hat{x}'^{L} \partial_{L}\frac{1}{r}
\label{STFProjection}\end{equation}
due to the identity
\begin{equation}
  \partial_L \frac{1}{r}
  =
  (-1)^l \frac{(2l - 1)!!}{r^{l+1}} \hat{n}_L; \quad L = i_1, \cdots i_l
\label{identity}\end{equation}
From this relation, one can systematically construct the STF tensor products $\hat{x}^{L}$ in the definitions of the mass and current multipole moments, \eqref{MLdef} and \eqref{currentmultipole} from \eqref{solPoisson} and \eqref{taylorExpansion}. These terms are all of first order in $G$, as also indicated by the subscript. There is no mass dipole term $M_a$ in the expansion due to the conventional choice of frame with origin at the center of mass.

For all orders in terms of mass multipole, we can write the expansion,
\begin{equation}\label{MPMexpansion}
\begin{split}
  h^{00}_{(1)}
  &=
  4\sum_{l=0}^{\infty} \frac{(2l-1)!!}{l !} \frac{\hat{n}^{L}M_{L}}{r^{l+1}}\,,
  \\
  h^{0i}_{(1)}
  &=
  4\sum_{l=1}^{\infty} \frac{l(2l-1)!!}{(l+1)!} \epsilon_{iab} \frac{\hat{n}^{a(L-1)} S_{b(L-1)}}{r^{l+1}}\,.
\end{split}
\end{equation}%
Truncating up to the mass quadrupole term $l = 0,2$, we have
\begin{equation}\label{intcond1}
  h_{(1)}^{00}\big|_{l=0,2}
  =
  \frac{4M}{r} + \frac{6M_{ab} }{r^3} \hat{n}_{ab} 
\end{equation}
Including only the leading spin term, the dipole ($l = 1$) we have
\begin{equation}\label{intcond2}
    \begin{split}
     h^{0i}_{(1)}\big|_{l=1} = -2\epsilon_{iab} n^a \frac{S^b}{r^2} ,
    \end{split}
\end{equation}
together with $h_{(1)}^{ij} = 0$.
If we are interested in a truncation at this order in the 1PM multipole expansion, the terms of Eq.~\eqref{intcond1} and \eqref{intcond2} will serve as initial conditions for our recursion relations that will 
generate the higher-order corrections in $G$ involving just these
three multipoles.

\section{Integrals}

In this appendix, we provide explicit evaluations of the generalized tensor bubble and Fourier integrals required for the main text.

\subsection{Tensor Bubble Integrals}\label{nbubble}
We define an arbitrary $d$-dimensional tensor bubble integral by
\begin{equation}
  \mathcal{B}^{(d),i_{1} \cdots i_{n}}_{n_1,n_2} (\boldsymbol\ell)
  =
  \int \frac{\mathrm{d}^d \boldsymbol{k}}{(2\pi)^d}
  \frac{k^{i_{1}} k^{i_{2}} \cdots k^{i_n}}{|\boldsymbol{k}|^{n_1} |\boldsymbol\ell - \boldsymbol{k}|^{n_2}}\,.
\end{equation}
This integral can be evaluated explicitly to give
\begin{equation}
  \mathcal{B}^{(d),i_{1} \cdots i_{n}}_{n_1,n_2} (\boldsymbol\ell)
  =
  \sum_{r=0}^{\lfloor n/2\rfloor}
  C^{d,n,r}_{n_{1},n_{2}} \frac{T_{r|\boldsymbol{\ell}}^{i_{1} \cdots i_{n}}}{|\boldsymbol{\ell}|^{n_{12}-d-2r}} \,,
\label{bubbleIntegral}\end{equation}
where $n_{12} = n_{1}+n_{2}$ and
\begin{equation}
\begin{aligned}
  &C^{d,n,r}_{n_{1},n_{2}}
  =
  \frac{\Gamma\big(\frac{n_{12}-d}{2}{-}r\big) \Gamma\big(n{-}r{+}\frac{d-n_1}{2}\big) \Gamma\big(r{+}\frac{d-n_2}{2}\big)}{(4\pi)^{\frac{d}{2}} 2^r \Gamma(\frac{n_1}{2})\Gamma(\frac{n_2}{2}) \Gamma\big(n+ d - \frac{n_{12}}{2}\big)} \,,
  \\
  &T_{r|\boldsymbol{\ell}}^{i_{1} \cdots i_{n}}
  =
  \frac{n! \delta^{(i_{1} i_{2}} \cdots \delta^{i_{2r-1} i_{2r}} \ell^{i_{2r+1}}\cdots \ell^{i_{n})}}{(2r)!(n-2r)!} \,.
\end{aligned}\label{BubbleCoeff}
\end{equation}
For the special case of $d=3$ and $n_{1}=n_{2}=2$, we define $\mathcal{I}^{i_{1}i_{2}\cdots i_{n}}_{\boldsymbol{\ell}} = \mathcal{B}^{(3),i_{1} \cdots i_{n}}_{2,2} (\ell)$. Its explicit form is
\begin{equation}
\begin{aligned}
  \mathcal{I}^{i_{1}i_{2}\cdots i_{n}}_{\boldsymbol{\ell}}
  &=
  \int \frac{d^{3} k}{(2\pi)^{3}} \frac{k^{i_{1}} k^{i_{2}} \cdots k^{i_n}}{|\boldsymbol{k}|^{2} |\boldsymbol{\ell} - \boldsymbol{k}|^{2}}\,,
  \\
  &=
  \sum_{r=0}^{\lfloor n / 2\rfloor}
  C_{n,r} T_{r|\boldsymbol{\ell}}^{i_{1}i_{2}\cdots i_{n}}|\boldsymbol{\ell}|^{2 r-1} \,.
\end{aligned}\label{2PMBubble}
\end{equation}
where
\begin{equation}
  C_{n,r} = \frac{(-1)^{r}\Gamma\big(n-r+\frac{1}{2}\big)}{2^{r+3} \sqrt{\pi}n!} \,,
\label{}\end{equation}
and $T_{r|\boldsymbol{\ell}}^{i_{1}i_{2}\cdots i_{n}}$ is as defined in Eq.~\eqref{BubbleCoeff}.

\subsection{Fourier transforms of tensor integrands}\label{FourierFormulaPart}

We now consider the Fourier transform in $d$-dimensions,
\begin{equation}
\begin{aligned}
  &\int \frac{d^{d} k}{(2 \pi)^{d}} 
  \frac{k^{i_{1}} \cdots k^{i_{m}}}{|\boldsymbol{k}|^{n}} e^{i k \cdot \boldsymbol{x}}
  \\
  &=
  \sum_{q=0}^{\lfloor m / 2\rfloor} 
  i^{m-2r} \frac{\Gamma\big(\frac{d-n+2m-2r}{2}\big)}{2^{n-m+r} \pi^{\frac{d}{2}}\Gamma(\frac{n}{2})} \frac{T_{r}^{i_{1} \cdots i_{m}}(\boldsymbol{x})}{r^{d-n+2 m-2 r}}\,,
\end{aligned}\label{FourierFormula}
\end{equation}
where 
\begin{equation}
  T_{r}^{i_{1} \cdots i_{m}}(x) 
  =
  \frac{m!\delta^{(i_{1} i_{2}} \cdots \delta^{i_{2r-1} i_{2 r}} \boldsymbol{x}^{i_{2 r+1}} \cdots \boldsymbol{x}^{i_{m})}}{(2r)!(m-2r)!}\,.
\label{}\end{equation}
Interestingly, this coincides with the definition $T^{N}_{r|\ell}$ in Eq.~\eqref{BubbleCoeff} if we replace $x$ by $\ell$, and $N = i_1 \cdots i_m$. The tensor
structure is thus identical to that of the bubble integrals.


\section{Axisymmetric stationary source}\label{fixedmultipole}

Consider the general definition of mass multipoles given in Eq.~\eqref{MLdef} for the case of a source with symmetry axis chosen as the $z$-axis so that the multipoles  $M^L$ are rank-$l$ STF tensors invariant under the $\mathrm{SO}(2)$ subgroup of $z$-axis rotations. In the far zone, the gravitational field is a scalar function on the sphere (angular dependence). Recall the linear gravitational field generated by $M_L$ in Eq.~\eqref{h1sol},
\begin{equation}
  h_{(1)}^{00}(\boldsymbol{x}) 
  =
  \sum_{l=0}^{\infty} \frac{4}{r^{l+1}} \frac{(2l - 1)!!}{l!} \,M_L \hat{n}^L \,.
\end{equation}
Let us define $\Phi_l(\hat{n})$ as
\begin{equation}
    h_{(1)}^{00}(\boldsymbol{x}) = \sum_{l=0}^{\infty} \frac{1}{r^{l+1}} \Phi_l (\hat{n}) \,.
\end{equation}
$\Phi_l(\hat{n})$ can be written in terms of the spherical harmonics. Imposing axisymmetry $\Phi_l(\theta,\phi)= \Phi_l(\theta)$ and using the orthonormality of the spherical harmonics, we can show that
\begin{equation}
\Phi_l(\theta,\phi)= 4 a_l\, \sqrt{\frac{4 \pi}{2l+1}} Y_{l 0}(\theta,\phi) = 4 a_l\, P_l(\cos\theta) \,,
\end{equation}
where $a_l$ is a number determined by $M_L$. A standard identity shown in Eq.~\eqref{poly} relates Legendre polynomials to STF products as,
\begin{equation}
  P_l(\cos\theta)
  =
  \frac{(2l-1)!!}{l!}\ \hat{n}^{L}\ \hat{s}_{L}\,, \qquad \cos\theta = \hat{n}\cdot\hat{s}.
\end{equation}
Thus,
\begin{equation}
  M_L\,\hat{n}^{L}
  =
  a_l\, \hat{n}^L \hat{s}_{L}.
\end{equation}
As $M_L$ are rank-$l$ STF tensors and recalling that STF tensors are nothing but alternative descriptions of the spherical harmoniocs, we can show that any angular dependence in $M_L$ should be solely possessed by $\hat{s}_{L}$
\begin{equation}
  M_L = a_l\, \hat{s}_{L}.
\label{eq:M-propto-zhat}\end{equation}
For an axisymmetric stationary source, each STF rank-$l$ multipole $M_L$ is fully characterized by a single number $a_l$. Therefore, for each $l$ the axisymmetric subspace is one-dimensional.

To illustrate, for the choice of spin axis $\hat{s}^i=\delta^{i3}$ the $l = 4$ STF tensor can explicitly be written in terms of $M_{3333}$ using the STF identity given in Eq.~\eqref{rank4stf}. If one imposes axisymmetry: $M_{ijkl} = a_4\, \hat{s}_{ijkl}$, then using $s^i = (0,0,1)$, we compute the independent components.

\paragraph{All--3 component}
\begin{align}  \hat{s}_{3333}
  &= 1 - \frac{6}{7} + \frac{3}{35} = \frac{8}{35}\,,
  \\
  M_{3333} &= a_4 \frac{8}{35} \implies a_4 = \frac{35}{8} M_{3333}\,.
\end{align}
\paragraph{Mixed components}
\begin{equation}
\begin{split}
  \hat{s}_{1133} &= \hat{s}_{2233} = 0 - \frac{1}{7} + \frac{1}{35} = -\frac{4}{35}\,,
  \\
  M_{1133} &= M_{2233} = a_4\!\left(-\frac{4}{35}\right) = -\frac{1}{2} M_{3333}\,.
\end{split}
\end{equation}
\paragraph{Pure transverse components}
\begin{equation}\begin{split}
  \hat{s}_{1111} &= \hat{s}_{2222} = \frac{3}{35}\,,
  \qquad
  \\
  M_{1111} &= M_{2222} = a_4 \frac{3}{35} = \frac{3}{8} M_{3333}\,.\end{split}
\end{equation}
\paragraph{Mixed transverse component}
\begin{equation}
  \hat{s}_{1122} = \frac{1}{35},
  \quad
  M_{1122}= a_4 \frac{1}{35} = \frac{1}{8} M_{3333}.
\end{equation}
\paragraph{Mixed odd component}
\vspace{2ex}
 $M_L$ being a Cartesian tensor invariant under rotations about the $z$--axis implies
\begin{align}
    M_{i_1\cdots i_l} = 0,
\end{align}
whenever the number of indices $ i_k \in \{x,y\}$ is odd. That means,
\begin{equation}
   M_{1233} = -    M_{1233} \implies M_{1233}  = 0.
\end{equation}
All remaining components follow from symmetry, index permutations, and tracelessness, $M_{iikl} = 0, \, \text{for all} \, k,l$ 
(e.g.\ $M_{1313}=M_{1133}$, etc.). Thus, for $l=4$, the single component $M_{3333}$ completely determines the axisymmetric STF tensor.

\section{2PM Kerr metric from multipoles}\label{kerrdecom}
In this section, we list the non-trivial components of the Kerr metric at 2PM order with expansion truncated at  $\mathcal{O}(a^4)$. Our
notation follows that of the main text where we introduced labels $Q[n]$ to indicate the origin of the different terms of the Kerr metric
based on the multipole expansion. 
\par
We list the 2PM-accurate components of the Kerr metric up to order $a^4$ as expressed in terms of the multipole labels $Q[n]$. For a concise expression, we introduce $x^{a}=\{x,y\}$ and $\tilde{x}^{a}=\{y,x\}$, where $a,b=1,2$.
\begin{enumerate}
\item[i.] \textbf{Diagonal components $h^{ii}_{(2)}\big|_{\mathcal O(a^4)}$:}
{\small
\begin{equation*}
\begin{split}
  & M^{-2}h^{aa}_{(2)}\big|_{\mathcal O(a^4)}
  \\
  &=\frac{ x_{a}^2}{r^{4}} Q[0]^2+ \frac{x_{a}^2(2r^2-15z^2)+r^2 \tilde{x}_{a}^2}{2r^8} Q[0]Q[2] 
  \\&
  -\frac{x_{a}^2(2r^2-9z^2)-r^2 \tilde{x}_{a}^2}{2r^8}Q[1]^2
  \\&
  +\bigg[
       \frac{3}{4r^6}{-}\frac{\tilde{x}_{a}^2{+}59z^2}{r^{8}}
    {+}\frac{7(4z^4{+}3\tilde{x}_{a}^2 z^2)}{2r^{10}}
    {+}\frac{105x_{a}^2 z^4}{4r^{12}}
  \bigg] Q[0]Q[4] 
  \\&
  - \bigg[
       \frac{3}{2r^6}
    {-}\frac{4\tilde{x}_{a}^2{+}47z^2}{2r^{8}}
    {+}\frac{2(11z^4{+}12\tilde{x}_{a}^2 z^2)}{r^{10}}
    {+}\frac{35x_{a}^2 z^4}{r^{12}} \bigg] Q[1]Q[3]
  \\&
  + \bigg[\frac{3x_{a}^2 }{r^{8}} + \frac{3z^2(\tilde{x}_{a}^2-8x_{a}^2)}{2r^{10}}
  + \frac{75x_{a}^2z^4}{4r^{12}} \bigg]Q[2]^2\,,
\end{split}
\end{equation*}
\begin{equation*}
\begin{split}
  &M^{-2}h^{33}_{(2)}\big|_{\mathcal O(a^4)} 
  \\&
  = \frac{ z^2}{r^{4}} Q[0]^2-\frac{(r^4-12r^2 z^2+15z^4)}{2r^8} Q[0]Q[2]\\& - \frac{(r^4-8r^2 z^2+9z^4)}{2r^8}Q[1]^2
  \\& -\frac{1}{4r^{12}}(2r^6-45r^4z^2+140r^2z^4-105z^6)Q[0]Q[4]
    \\
    & +\frac{1}{r^{12}}(r^6-18r^4z^2+50r^2z^4-35z^6)Q[1]Q[3]
    \\
    & -\frac{1}{4r^{12}}(2r^6-39r^4z^4+108r^2z^4-75z^6)Q[2]^2 
    \end{split}
\end{equation*}
\item[ii.] \textbf{Off-Diagonal components ($( h^{ij}_{(2)}\big|^{(i \neq j )}_{\mathcal O(a^4)})$):}
\begin{equation*}
\begin{split}
 M^{-2} h^{12}_{(2)}\big|_{\mathcal O(a^4)}
  &=
  \frac{ x y}{r^{12}} \Big[
    4r^{8}Q[0]^2
    + 2r^{4}(r^2-15z^2) Q[0]Q[2] 
    \\&\quad
    - 6r^{4}(r^2-3z^2)Q[1]^2
    \\&\quad
    +4(r^4-42 r^2 z^2 +105 z^4)Q[0]Q[4] 
    \\&\quad
    -4(2r^4-24 r^2 z^2 +35 z^4)Q[1]Q[3]
  \\& 
    +3(r^4-18r^z z^2 +25z^4)Q[2]^2 
  \Big]\,,\\
  M^{-2} h^{a3}_{(2)}\big|_{\mathcal O(a^4)}
  &= 
  \Big[4r^{8} Q[0]^2
  +2 (7r^2-15z^2)r^{4} Q[0]Q[2] 
  \\&\quad
  -2(5r^2-9z^2)r^4 Q[1]^2
  \\&\quad
  +(17r^4-98r^2z^2+105z^4) Q[0]Q[4]
  \\&\quad
  -4(7r^4-36r^2z^2+35z^4)Q[1]Q[3]
  \\&\quad
  +3(5r^4-26r^2z^2+25z^4)Q[2]^2\Big] \frac{x_{a}z}{4r^{12}} \,,
\end{split}\label{}
\end{equation*}
\item[iii.] 
\textbf{Current components $( h^{0i}_{(2)}\big|_{\mathcal O(a^3)})$:}
\begin{equation*}
\begin{split}
  & M^{-2} h^{0a}_{(2)}\big|_{\mathcal O(a^3)}\\
  &=
   \Big[ 3 (r^2-5z^2) Q[0]Q[3]
    +(r^2+3z^2)Q[1]Q[2]\Big]\frac{\epsilon_{ab}x^{b}}{2r^8}
\end{split}
\end{equation*}
}
\end{enumerate}

~\\~\\~\\~\\~\\~\\~\\~\\~\\~\\~\\~\\~\\~\\~\\~\\~\\~\\~\\~\\~\\~\\~\\~\\

\newpage


\end{document}